\documentclass{article}
\usepackage[a4paper,top=2.54cm,bottom=2.54cm,left=3.17cm,right=3.17cm,%
            includehead,includefoot]{geometry}

\usepackage{amsmath,amssymb,amsfonts,amsthm}
\usepackage{graphicx}
\usepackage{float}
\usepackage{xcolor}
\usepackage[numbers,square,sort&compress]{natbib}
\usepackage{hyperref}
\hypersetup{colorlinks,citecolor=blue,linkcolor=blue,breaklinks=true}
\usepackage{epstopdf}
\usepackage{iitem}
\usepackage{natbib}
\usepackage{amsmath}
\usepackage[ruled,vlined]{algorithm2e}
\usepackage{multirow}
\usepackage{verbatim}
\usepackage{authblk}
\usepackage{yhmath} 

\allowdisplaybreaks

\begin{document}
\title{DRVN (Deep Random Vortex Network): A new physics-informed machine learning method for simulating and inferring incompressible fluid flows}
\author[b,d]{Rui Zhang\thanks{This work was done when the first and the second authors were visiting Microsoft Research.}}
\author[b,d]{Peiyan Hu}
\author[a]{Qi Meng \thanks{E-mails: rayzhang@amss.ac.cn, meq@microsoft.com}}
\author[a]{Yue Wang}
\author[c]{Rongchan Zhu}
\author[b]{Bingguang Chen}
\author[b]{Zhi-Ming Ma}
\author[a]{Tie-Yan Liu}

\affil[a]{Microsoft Research, Danling Street, Haidian, Beijing, China}
\affil[b]{Academy of Mathematics and Systems Science, Chinese Academy of Sciences, Zhongguancun East Road, Beijing, China}
\affil[c]{Bielefeld University, Bielefeld, North Rhine-Westphalia, Germany}
\affil[d]{University of Chinese Academy of Sciences, No.19 Yuquan Road, Beijing, China}

\maketitle

\begin{abstract}
We present the \emph{deep random vortex network} (DRVN), a novel physics-informed framework for simulating and inferring the fluid dynamics governed by the incompressible Navier--Stokes equations. Unlike the existing physics-informed neural network (PINN), which embeds physical and geometry information through the residual of equations and boundary data, DRVN automatically embeds this information into neural networks through neural random vortex dynamics equivalent to the Navier--Stokes equation. Specifically, the neural random vortex dynamics motivates a Monte Carlo-based loss function for training neural networks, which avoids the calculation of derivatives through auto-differentiation. Therefore, DRVN can efficiently solve Navier--Stokes equations with non-differentiable initial conditions and fractional operators. Furthermore, DRVN naturally embeds the boundary conditions into the kernel function of the neural random vortex dynamics and thus, does not need additional data to obtain boundary information. We conduct experiments on forward and inverse problems with incompressible Navier--Stokes equations. The proposed method achieves accurate results when simulating and when inferring Navier--Stokes equations. For situations that include singular initial conditions and agnostic boundary data, DRVN significantly outperforms the existing PINN method. Furthermore, compared with the conventional adjoint method when solving inverse problems, DRVN achieves a 2 orders of magnitude improvement for the training time with significantly precise estimates.
\end{abstract}

\maketitle
\section{Introduction}\label{sec1}
The ubiquitous Navier--Stokes equations (NSEs) are indispensable for modeling fluids ranging from those studied in meteorology to ocean currents. \cite{batchelor2000introduction, smits2000physical, temam2001navier, doi:10.1063/5.0087550, WANG2021103223} Consequently, since NSEs are widespread in scientific, \cite{jaccod2021constrained, doi:10.1146/annurev-fluid-010719-060317} industrial, \cite{jameson1998optimum, doi:10.1146/annurev-fluid-122316-044531} and engineering applications, \cite{Reinfoce,SRIRAM2021102883} gaining a deep understanding of them is important. Recently, with the growth of computer technology and better availability of data, the development of deep learning techniques has led to many advances. Deep learning methods, such as physics-informed neural networks (PINNs), can help to solve partial differential equations (PDEs) accurately and efficiently, \cite{ raissi2019physics, doi:10.1126/science.aaw4741, jin2021nsfnets, cai2022physics, han2018solving, long2018pde, doi:10.1063/5.0038929} which demonstrates the promising future of combining deep learning with scientific computing. In general, PINNs seamlessly embed information relating to the observed data and physical laws into neural networks via an automatic differentiation regime. The loss function of a PINN contains the supervised loss constructed from the observed data and the physical loss of PDEs, including the residuals of the governing equations and other additional conservation laws. Compared with conventional numerical methods, PINNs are mesh-free and can handle inverse problems efficiently due to their differentiability. \cite{raissi2019physics}

Recently, significant scientific effort has been devoted to utilizing PINNs to solve scientific problems based on the NSEs. For instance, \cite{doi:10.1126/science.aaw4741} developed hidden fluid mechanics to solve forward and inverse fluid mechanics problems in arbitrarily complex domains. \cite{SUN2020161} combined PINNs with the Bayesian method to reconstruct flow fields from sparse and noisy fluid data. \cite{jin2021nsfnets} proposed Navier--Stokes flow nets (NSFnets) by considering the velocity--pressure and the vorticity--velocity formulations simultaneously to simulate both laminar and turbulent flows. Moreover, several methods have been proposed to optimize the network architectures and training dynamics of PINNs, e.g.,  multi-scale deep neural networks, \cite{CiCP-28-1970} hard-constraint PINNs, \cite{lu2021physics} and the dynamic pulling method. \cite{kim2021dpm}

While these approaches have made remarkable progress, using them to solve the NSEs faces the following three fundamental challenges. First, the existing frameworks of PINNs embed the physical information of fluids via the residual of the NSEs, i.e., by utilizing derivative information directly in PDEs to define the loss function. This implicitly assumes that the relevant functions are sufficiently smooth, which limits their ability to deal with non-differentiable functions, e.g., the Dirac delta function. The second challenge relates to their efficiency when calculating high-order or fractional-order derivatives. For example, fractional NSEs have been widely adopted in modeling fluid dynamics involving historical memory and long-range interactions. \cite{zhou2017time,zhou2017weak, cholewa2018fractional, doi:10.1063/1.5123118} However, there is a high computational cost in estimating the fractional operator. \cite{herrmann2011fractional, lai2016investigation} 
The third challenge relates to the inefficiency of tuning the hyperparameters for PINNs. \cite{krishnapriyan2021characterizing, wang2021understanding, PSAROS2022111121} PINNs use a weighted summation of the residual of the equation and the residual of the boundary and initial conditions, where the performance is sensitive to the weight. Training a PINN generally requires boundary data over the whole period to depict the boundary information, \cite{doi:10.1126/science.aaw4741, jin2021nsfnets} which may be unavailable or redundant.  Therefore, concisely integrating machine learning and the physical information about the fluid flow is also critical.

In this work, we address these challenges by combining deep learning with a reformulation of the random vortex method (RVM), \cite{chorin1973numerical, long1988convergence, majda_bertozzi_2001,doi:10.1063/5.0065073} which is a mesh-free algorithm for implementing the fluid mechanics equations. Instead of solving the NSEs directly, RVM converts the velocity field to its corresponding probabilistic representation via the Feynman--Kac formula, which can be approximated by a Monte Carlo method. In the RVM formulation, the spatial derivation in the original formulation of an NSE can be approximated by sampling from a stochastic differential equation (SDE) driven by a L\'evy process. In this way, the RVM can efficiently handle non-smooth and fractional equations. Therefore, we propose a novel physics-informed machine learning framework, namely the deep random vortex network (DRVN), which utilizes a deep neural network to represent the velocity field. The loss function is constructed according to its probabilistic representation in RVM. There are four attractive advantages of DRVN:
\begin{enumerate}
	\item Broad range of applications. Compared with the existing PINN framework, DRVN can more easily handle non-smooth and fractional equations. DRVN requires only that the network function is integrable in the domain, rather than the second-order or fractional-order continuously differentiability necessary for PINN. Thus, it can represent non-smooth solutions and initial conditions. Furthermore, calculating fractional derivatives can be replaced via the efficient sampling from the L\'evy process, which does not increase the algorithmic complexity.
  
  \item Boundary data-free. Unlike existing PINNs, which require data points on the boundary during the whole period to embed geometrical information into their neural networks, DRVN utilizes the kernel function as prior knowledge to implicitly constrain the neural networks so that they satisfy the boundary conditions. Thus, DRVN does not need additional boundary data and can handle situations where boundary data are unavailable or expensive.

  \item Easy to implement. Instead of the loss function in a PINN, which is constituted by the equation term, boundary condition term, data term, and other information, all the physical and geometrical information is naturally embedded into the formulation of RVM, so that, in general, the loss function of DRVN has only one term. Thus, there are fewer hyperparameters in the DRVN loss function, which saves effort in fine-tuning the hyperparameters.
  
  \item Efficient for the inverse problem. Compared with the classical RVM, DRVN constructs a continuous model that directly maps the spatial-temporal coordinates to the velocity. By leveraging the auto-differentiation of deep neural networks, DRVN achieves fast inference compared to the traditional inference algorithm based on RVM, i.e., the adjoint method.
\end{enumerate}

We demonstrate the effectiveness of DRVN by solving forward and inverse problems for various equations, including 2-dimensional (2D) and 3-dimensional (3D) Lamb--Oseen vortices (with a singular initial condition), 2D fractional NSEs, and a 2D Taylor--Green vortex (with a periodic boundary condition). For forward problems, the relative $\ell_2$ errors are around $1\%$ for most equations. For situations with singular initial conditions and agnostic boundary data, DRVN significantly outperforms the existing PINN method.  For inverse problems, we utilize a parametric solver to infer the viscosity term $\nu$ in the 2D NSE and the diffusion parameter $\alpha$ in the fractional NSE. Compared with the traditional adjoint method, DRVN achieves 2 orders of magnitude improvement for the training time with significantly precise estimates.

This paper is organized as follows. In Section~\ref{sec2}, we describe the notation and the problem setups. In Section~\ref{sec3}, we introduce our methodology using the 2D NSEs as an example. In Section~\ref{sec4}, we report the results of numerical experiments that demonstrate the effectiveness of DRVN. Finally, we summarize and discuss our method in Section~\ref{sec5}.

\section{Notation and Problem Setups}\label{sec2}
\subsection{Notation}
In this section, we introduce the notation used in this paper. We utilize bold letters for vectors and matrices. For a matrix $\boldsymbol{A}$, $\boldsymbol{A}_{i, j}$ denotes its $(i, j)$th entry. For a vector $\boldsymbol{a}$, $\boldsymbol{a}_i$ and $\|\boldsymbol{a}\|_{2}$ are its $i$th entry and Euclidean norm, respectively. For a 2D vector $\boldsymbol{a}$, then its orthogonal complement $\boldsymbol{a}^{\perp}:=(-\boldsymbol{a}_2, \boldsymbol{a}_1)$. $\boldsymbol{I}_n$ is an $n\times n$ identity matrix. We use $\langle\cdot,\cdot\rangle$ for the standard Euclidean inner product between two vectors. 

Dirac's delta function is 
\begin{equation}
\delta(x) = \begin{cases}
      +\infty, & x=0, \\
      0,       & x \neq 0.
\end{cases}
\end{equation}
$\delta(\cdot)$ satisfies that $\int_{-\infty}^{+\infty} \delta(x)\psi(x) \mathrm{d} x=\psi(0)$ for all smoothing test functions $\psi(\cdot)$. 
We utilize $\lfloor\cdot\rfloor:\mathbb{R} \to \mathbb{Z}$ to denote the greatest integer.
The velocity and vorticity terms of an NSE are $\boldsymbol{u}$ and $\boldsymbol{\omega}$, respectively. 
Furthermore, $(u,v)$ is the velocity $\boldsymbol{u}$ in $\mathbb{R}^2$. Re and $\nu=1/Re$ are the Reynolds number and the viscosity. $\Omega$ and $\partial \Omega$ denote the domain and boundary of the equations, respectively.

\subsection{Problem setups}\label{sec2.2}
We consider 2D incompressible NSEs defined in the domain $\Omega\in\mathbb{R}^2$:
\begin{equation}\label{nse_v}
    \begin{aligned}
    \frac{\partial \boldsymbol{u}}{\partial t} + (\boldsymbol{u}\cdot\nabla) \boldsymbol{u}  &= \nu \Delta \boldsymbol{u} -\nabla p, \quad \text{in}\ \Omega,\\
   \nabla\cdot \boldsymbol{u} &=0,
\end{aligned}
\end{equation}
where $\boldsymbol{u}(\boldsymbol{x},t) \in \mathbb{R}^2$ is the velocity field, $p$ is the pressure term, and $\nu > 0$ is the viscosity. The vorticity is $\omega=\nabla \times \boldsymbol{u}\in \mathbb{R}$, which evolves according to the following vorticity equation:
\begin{equation}\label{nse_omega}
\begin{aligned}
\frac{\partial \omega}{\partial t}&=-(\boldsymbol{u}\cdot \nabla)\omega + \nu\Delta \omega, \quad \text{in}\ \Omega,\\
\omega &=\nabla\times \boldsymbol{u}.
\end{aligned} 
\end{equation}where $\boldsymbol{u}(\boldsymbol{x},t) \in \mathbb{R}^2$ is the velocity field, $\omega(\boldsymbol{x},t)\in\mathbb{R}$ is the vorticity, and $\nu > 0$ is the viscosity. In this paper, the numerical method is designed based on the vorticity form of the equation.

The 2D fractional vorticity form of the NSE is given by the following equations:
\begin{equation}\label{frac_omega}
\begin{aligned}
\frac{\partial \omega}{\partial t}+( \boldsymbol{u}\cdot \nabla)\omega &=-\nu(-\Delta)^{\alpha/2} \omega, \quad\text{in} \ \mathbb{R}^2\\
\omega &=\nabla\times \boldsymbol{u},
\end{aligned}
\end{equation}
where the diffusion parameter $\alpha$ is restricted to the interval $(0, 2)$. Notice the fractional Laplacian $(-\Delta)^{\alpha/2}$ is on the right-hand side of Eq.~(\ref{frac_omega}), which is defined by directional derivatives. \cite{lischke2020fractional, pang2019fpinns}

Our aim is to simulate and infer the fluid flow based on the NSE, which are described as the forward problem and inverse problem, respectively:
\begin{itemize}
    \item \textbf{Forward problem: }Given the initial velocity field $\boldsymbol{u}(\boldsymbol{x},0)$ and vorticity field $\omega(\boldsymbol{x},0)$, the general forward problem aims to simulate the velocity field in the time--space domain $\Omega \times [0,T]$ for a given viscosity term $\nu$ or simultaneously for a set of viscosity terms $\nu$. 
    
    \item \textbf{Inverse problem: }Given the initial velocity field $\boldsymbol{u}(\boldsymbol{x},0)$, vorticity field $\omega(\boldsymbol{x},0)$, and the observable dataset $\mathcal{D}: \{\boldsymbol{x}^{(d)},t^{(d)}; \boldsymbol{u}^{(d)}\}_{d=1}^D$ generated from a system that satisfies the NSEs, the target of the inverse problem is to infer the unknown parameters of the system (e.g., the viscosity term $\nu$) from the observable data.
\end{itemize}

\section{Methodology of DRVN}\label{sec3}
In this section, we introduce DRVN, which utilizes a feedforward neural network (FNN) to parameterize the velocity field of the fluid flow and simulate the dynamics by optimizing a loss function based on the random vortex dynamics.  To efficiently approximate the Feynman--Kac formula in the random vortex dynamics, DRVN utilizes a two-phase Monte Carlo method to obtain an unbiased estimate. We will apply DRVN to simulate and infer an incompressible fluid governed by the NSEs. 

\subsection{Feedforward neural network}
An FNN is a parameterized continuous function $\boldsymbol{F}_{\text{NN}}(\cdot): \mathbb{R}^{d_1}\rightarrow\mathbb{R}^{d_{L+1}}$ that is constructed as:
\begin{align}
    \boldsymbol{F}_{\text{NN}}(\boldsymbol{x}; \boldsymbol{\Theta})=\theta^L\sigma(\theta^{L-1}\cdots \sigma(\theta^1\boldsymbol{x})),
\end{align}where $\boldsymbol{x}\in\mathbb{R}^{d_1}$ denotes the input, $\boldsymbol{\Theta}=[{\theta^1},\theta^2,\dots,\theta^L]$ where $\theta^l\in\mathbb{R}^{d_l\times d_{l+1}}$ denotes the parameter matrix at layer $l$, and $\sigma(\cdot)$ is an element-wise non-linear transformation called the activation function. There are two widely used activation functions:
\begin{equation}
    \begin{aligned}
        \text{ReLU} (z) &= \max (z,0),\\
        \tanh(z) &= \frac{e^{z}-e^{-z}}{e^{z}+e^{-z}}.
    \end{aligned}
\end{equation}
An FNN is a powerful function approximator in the continuous function class.\cite{hornik1989multilayer} In the following, we will use an FNN to parameterize the velocity field of the fluid flow.

\subsection{Neural random vortex dynamics}\label{sec3.1}
In this section, we introduce the neural random vortex dynamics. The vorticity evolves according to the parabolic form of Eq.~(\ref{nse_omega}). The Feynman--Kac equation links the vorticity $\omega(\boldsymbol{x},t)$ with the following stochastic process:
\begin{equation}\label{X_dynamic}
    d\boldsymbol{X}_t=\boldsymbol{u}(\boldsymbol{X}_t,t)dt+\sqrt{2\nu}d\boldsymbol{B}_t, \qquad \boldsymbol{X}_0=\boldsymbol{\xi}, 
\end{equation}
where $\boldsymbol{\xi}\in\Omega$ represents the initial spatial coordinate in $\Omega$, which is sampled from the initial vorticity distribution $\omega_0$, $\boldsymbol{B}_t$ is the 2D Brownian motion, and $\boldsymbol{X}_t$ is a diffusion process that satisfies Taylor's formulation of Brownian motion. \cite{taylor1922diffusion, long1988convergence} It has been proved that the probability density of $\boldsymbol{X}_t$ follows $\omega(x,t)$. \cite{chorin1973numerical, long1988convergence} From the relations $\omega=\nabla \times \boldsymbol{u}$ and $\nabla\boldsymbol{u}=0$ in $\Omega$, the velocity can be represented as $\boldsymbol{u}(\boldsymbol{x},t)=\int_{\Omega}\boldsymbol{K}(\boldsymbol{x}-\boldsymbol{y})\omega(\boldsymbol{y},t)d\boldsymbol{y}$, where the kernel $\boldsymbol{K}(\cdot)$ is determined by the boundary condition.
Then, the velocity field has the following representation: \cite{chorin1973numerical, long1988convergence, goodman1987convergence, long1988convergence}
\begin{equation}\label{pro_rep}
     \boldsymbol{u}(\boldsymbol{x},t) = \int_{\Omega}\mathbb{E}[\boldsymbol{K}(\boldsymbol{x}-\boldsymbol{X}_t(\boldsymbol{\xi}))]\omega(\boldsymbol{\xi},0)d\boldsymbol{\xi}.
\end{equation}

To simulate the velocity field governed by Eqs.~(\ref{nse_v}) and~(\ref{nse_omega}), we reformulate the stochastic process in Eq.~(\ref{X_dynamic}) by replacing $\boldsymbol{u}(\boldsymbol{X}_t,t)$ with a FNN $\boldsymbol{u}_{\text{NN}}(\boldsymbol{x},t; \boldsymbol{\Theta})$ with input $(\boldsymbol{x},t)$ and given parameter $\boldsymbol{\Theta}$:
\begin{equation}\label{X_dynamicF}
   d\boldsymbol{X}_t=\boldsymbol{u}_{\text{NN}}(\boldsymbol{X}_t,t; \boldsymbol{\Theta})dt+\sqrt{2\nu}d\boldsymbol{B}_t, \qquad \boldsymbol{X}_0=\boldsymbol{\xi}.
\end{equation}
We call the above process the neural random vortex dynamics. Then, our goal is to find the optimal value of parameter $\boldsymbol{\Theta}^*$ such that the neural network function is equal to the velocity $\boldsymbol{u}(\boldsymbol{x},t)$ in Eq.~(\ref{pro_rep}), i.e., $\boldsymbol{u}_{\text{NN}}(\boldsymbol{x},t; \boldsymbol{\Theta}^*)=\int_{\Omega}\mathbb{E}[\boldsymbol{K}(\boldsymbol{x}-\hat{\boldsymbol{X}}_t(\boldsymbol{\xi}))]\omega(\boldsymbol{\xi},0)d\boldsymbol{\xi}$, where $\hat{\boldsymbol{X}}_t(\boldsymbol{\xi})$ is sampled from Eq.~(\ref{pro_rep}) with $\boldsymbol{u}_{\text{NN}}(\boldsymbol{x},t; \boldsymbol{\Theta}^*)$.

Concretely, we find $\boldsymbol{\Theta}^*$ by solving the following optimization problem: 
\begin{equation}\label{opt_pro}
    \boldsymbol{\Theta^*} =  \operatorname{argmin}_{\Theta}\int_{\Omega\times[0,T]}
    \left\| \boldsymbol{u}_{\text{NN}}(\boldsymbol{x},t; \boldsymbol{\Theta}) 
    - \int_{\Omega}\mathbb{E}[\boldsymbol{K}(\boldsymbol{x}-\hat{\boldsymbol{X}}_t(\boldsymbol{\xi}))]\omega(\boldsymbol{\xi},0)d\boldsymbol{\xi} \right\|_2^2 d\boldsymbol{x}dt,
\end{equation}
where $\hat{\boldsymbol{X}}_t(\boldsymbol{\xi})$ is sampled from Eq.~(\ref{pro_rep}). 

\subsection{Algorithm with a Monte Carlo estimator}
In this section, we introduce the numerical method used to find $\Theta^*$. We discuss the temporal discretization, the architecture of the neural network, the Monte Carlo method for estimating the integrals, and the gradient-based search algorithm.  

We divide the time period $[0, T]$ into $M$ uniform intervals, i.e., $0=t_0< t_1<\cdots<t_M = T$. Given the vorticity field $\omega(\boldsymbol{\xi},0)$ for all coordinate points in $\Omega$ at $t=t_0$, we adopt the following Euler discretization of Taylor's formulation of Brownian motion in Eq.~(\ref{X_dynamic}) to calculate the path of ${\boldsymbol{X}_{t_m}}$ for all $\boldsymbol{\xi}$:
\begin{equation}\label{eluer}
    \boldsymbol{X}_{t_{m}}(\boldsymbol{\xi}) - \boldsymbol{X}_{t_{m-1}}(\boldsymbol{\xi}) = \boldsymbol{u}_{\text{NN}}( \boldsymbol{X}_{t_{m-1}}(\boldsymbol{\xi}); \boldsymbol{\Theta}_{m-1}) \Delta t + \sqrt{2 \nu} \Delta \boldsymbol{B}_{m},
\end{equation}
where $\Delta t  = t_{m} - t_{m-1}$ and $\Delta \boldsymbol{B}_m = \boldsymbol{B}_{t_{m}} - \boldsymbol{B}_{t_{m-1}}$. Note that we use $M$ subnetworks  $\boldsymbol{u}_{\text{NN}}(\boldsymbol{x}; \boldsymbol{\Theta}_{m})$, $m\in \{1, 2, \dots, M\}$, each of which has input $\boldsymbol{x}$ and parameter $\boldsymbol{\Theta}_{m}$.

\begin{figure*}[!htbp]
\centering
\centerline{\includegraphics[width=1.0\linewidth]{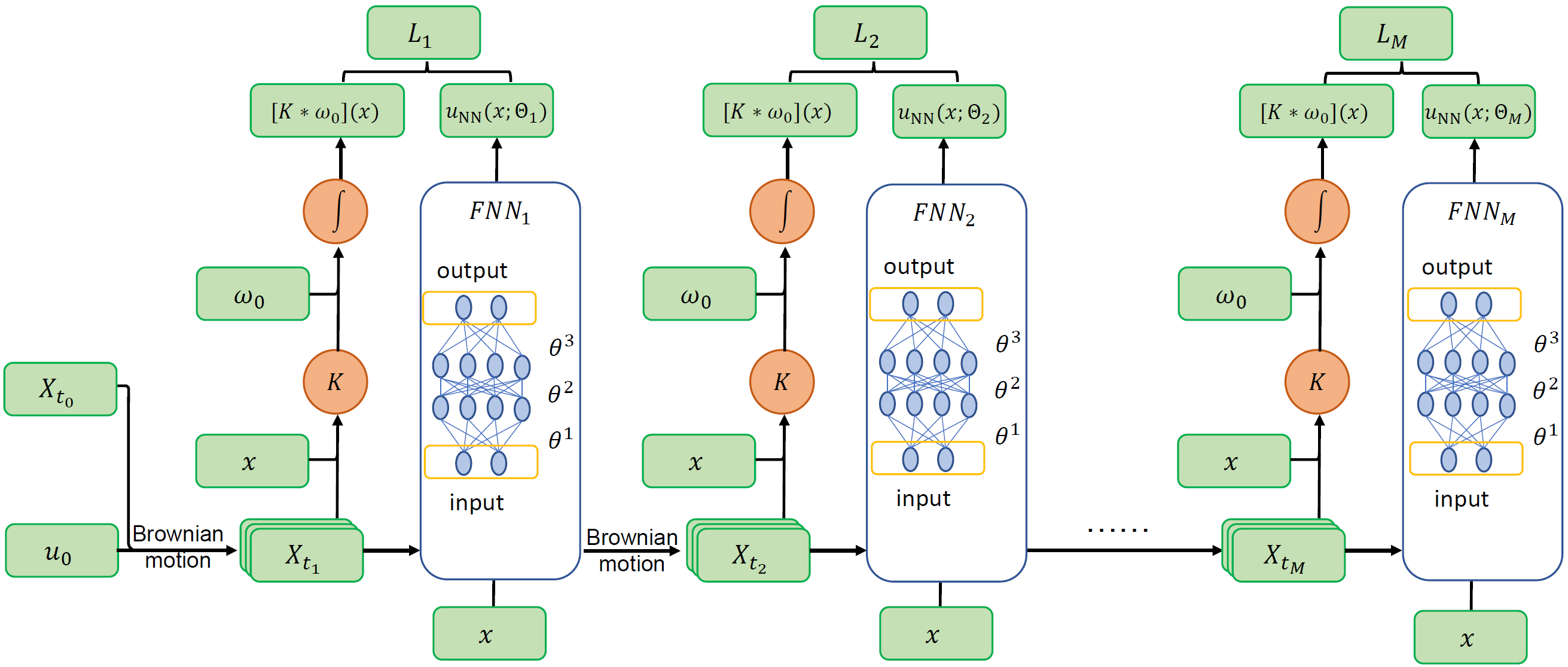}}
\caption{Illustration of DRVN for a 2D NSE. The total time is divided uniformly into $M$ intervals. Each column for $t_m$ corresponds to $\text{FNN}_m$, where $1\leq m \leq M$. Starting from input $\boldsymbol{x}$, then $\boldsymbol{u}_{\text{NN}}(x;\Theta_m)$ and  $G(\boldsymbol{x},t_m)$ are obtained via $\text{FNN}_m$ and the stochastic process in Eq.~(\ref{pro_rep}), respectively. The loss function is constructed according to Eq.~(\ref{loss}).}
\label{fig:framework}
\end{figure*}

We evaluate the $L_2$ norm on the right-hand side of Eq.~(\ref{opt_pro}) on uniformly sampled grid points and  
search for $\boldsymbol{\Theta}^*$ by optimizing the following loss function:
\begin{equation}\label{loss}
\mathcal{L}(\boldsymbol{\Theta}) = \sum_{b=1}^{B}\sum_{m=1}^{M} \left\| \boldsymbol{u}_{\text{NN}}(\boldsymbol{x}^{(b)};\boldsymbol{\Theta}_{m})- \int_{\Omega}\mathbb{E}[\boldsymbol{K}(\boldsymbol{x}^{(b)}-\hat{\boldsymbol{X}}_t(\boldsymbol{\xi}))]\omega(\boldsymbol{\xi},0)d\boldsymbol{\xi} \right\|_2^2, 
\end{equation}                  
where $B$ is the batch size per epoch and $\boldsymbol{\Theta} = \{\boldsymbol{\Theta}_{1}, \boldsymbol{\Theta}_{2}, \dots, \boldsymbol{\Theta}_{M} \}$ are the parameters in each subnetwork. Figure~\ref{fig:framework} illustrates the architecture of DRVN.

We utilize an unbiased Monte Carlo method to estimate the integrals and expectation in Eq.~(\ref{loss}). We sample initial coordinate points $\{\boldsymbol{\xi}^{(i)}\}_{i=1}^{I}$  distributed in $\Omega$ uniformly and the corresponding vorticity field $\{\omega(\boldsymbol{\xi}^{(i)},0)\}_{i=1}^{I}$ at $t=t_0$ to calculate the integral over $\omega$. Furthermore, to approximate the expectation of the kernel function in Eq.~(\ref{pro_rep}), we utilize a Monte Carlo method to sample $N$ paths independently for each $\boldsymbol{\xi}^{(i)}$ in the diffusion process Eq.~(\ref{eluer}), and denote these as $\{ \boldsymbol{X}^n_{t_{m}}(\boldsymbol{\xi}^{(i)})\}_{n=1}^{N}$ for all $m \in \{1,2,\dots, M\}$. Then, the term $\int_{\Omega}\mathbb{E}[\boldsymbol{K}(\boldsymbol{x}^{(b)}-\hat{\boldsymbol{X}}_t(\boldsymbol{\xi}))]\omega(\boldsymbol{\xi},0)d\boldsymbol{\xi}$ is estimated using a Monte Carlo method as
\begin{equation}\label{mc_u_hat}
\boldsymbol{G}(\boldsymbol{x},t_m):=\frac{|\Omega|}{I}\sum_{i=1}^{I}\sum_{n=1}^N \frac{1}{N} \boldsymbol{K}(\boldsymbol{x}-\boldsymbol{X}^n_{t_m}(\boldsymbol{\xi}^{(i)}))\omega(\boldsymbol{\xi}^{(i)},0),
\end{equation}
where $|\Omega|$ is the area of the domain $\Omega$. Notice that a two-phase Monte Carlo method is utilized in Eq.~(\ref{mc_u_hat}) to find unbiased estimates of the integration and expectation terms.

We use a gradient-based optimization algorithm to find $\boldsymbol{\Theta}^*$. A basic optimization algorithm widely used in deep learning is stochastic gradient descent:
\begin{align}
    \boldsymbol{\Theta}^k=\boldsymbol{\Theta}^{k-1}-\eta\cdot\nabla_{\boldsymbol{\Theta}}\mathcal{L}(\boldsymbol{\Theta}^{k-1}),
\end{align}
where $k$ is the iteration index. This algorithm starts from the initialization point $\boldsymbol{\Theta}^0$. Other variants, like Adam,\cite{kingma2014adam} are also widely used as optimizers to minimize the loss function in machine learning.

\subsection{Implementation details}
\subsubsection{Forward problems}
When solving forward problems for initial vortex $\omega_0$ and a specific $\nu$, DRVN approximates the velocity field via neural networks $\boldsymbol{u}_{\text{NN}}(\boldsymbol{x};\boldsymbol{\Theta}_m)$ and constructs the loss function via Eq.~(\ref{loss}). DRVN utilizes Adam \cite{kingma2014adam} to optimize the parameter $\boldsymbol{\Theta}$, which is a variant of stochastic gradient descent for training deep models. Algorithm~\ref{algo} illustrates the framework of DRVN for solving the 2D NSE forward problem.

Besides solving a specific NSE, we also applied DRVN to parametric solver learning, which aims to learn a generalizable model that can simultaneously output the velocity for a class of NSEs with different parameters. 
As the Reynolds number is directly related to the complexity of the numerical simulation and turbulence, developing a stable solver that is capable of solving NSEs with different values of $\nu$ is very important.  For the 2D NSE, we use DRVN to obtain a neural network function $\boldsymbol{u}_{\text{NN}}(\boldsymbol{x},\nu, t)$ that can be generalized to a range of $\nu$ in Eq.~(\ref{nse_v}). Thus, we regard $\nu$ as an input to the neural network, and we sample different values of $\nu$ and $\boldsymbol{x}\in \Omega$ synchronously during training. We consider the following loss function:
\begin{equation}\label{loss_ol}
    \mathcal{L}(\boldsymbol{\Theta}) = \sum_{p=1}^{P}\sum_{b=1}^{B}\sum_{m=1}^{M} \|\boldsymbol{u}_{\text{NN}}(\boldsymbol{x}^{(b)}, {\nu}_p;\boldsymbol{\Theta}_m)- \boldsymbol{G}(\boldsymbol{x}^{(b)}, {\nu}_p, t_m)\|_2^2,
\end{equation}
where $B$ and $P$ are the numbers of values of $\boldsymbol{x}$ and $\nu$ sampled in each epoch, respectively. Furthermore, we also learn a parametric solver for different values of the diffusion parameter $\alpha$ in our experiments with a 2D fractional NSE. The forward propagation, the construction of the loss function, and the optimization algorithm are all the same as when solving a forward problem.  

\begin{algorithm}[H]
\caption{2D Deep Random Vortex Network (DRVN)}
\label{algo}
\LinesNumbered 
\KwIn{Coordinates $\{\boldsymbol{\xi}^{(i)}\}_{i=1}^{I}$, initial vortex $\{w(\boldsymbol{\xi}^{(i)},0)\}_{i=1}^{I}$, neural network $\{\boldsymbol{u}_{\text{NN}}(\boldsymbol{x},t_m)\}_{m=1}^{M}$.}
Simultaneously initialize the parameter $\{\boldsymbol{\Theta}_{t_m}\}_{m=1}^M$ of the neural networks $\{\boldsymbol{u}_{\text{NN}}(\boldsymbol{x},t_m)\}_{m=1}^M$ via Xavier method\;
\For{$E$ epochs}{
Initialize  $\boldsymbol{X}_{t_0}(\boldsymbol{\xi}^{(i)}) = \boldsymbol{\xi}^{(i)}$\;  
Sample $\{\boldsymbol{x}^{(b)}\}_{b=1}^{B}$ uniformly in $\Omega$\;
$\mathcal{L}$ = 0\;
\For{$M$ steps}{
    $\boldsymbol{X}_{t_{m}}(\boldsymbol{\xi}^{(i)}) = \boldsymbol{X}_{t_{m-1}}(\boldsymbol{\xi}^{(i)}) + \boldsymbol{u}_{\text{NN}}( \boldsymbol{X}_{t_{m-1}}(\boldsymbol{\xi}^{(i)}), t_{m-1}) \Delta t + \sqrt{2 \nu} \Delta \boldsymbol{B}_{m}$\;
    $\boldsymbol{\hat{u}}_{\text{NN}}(\boldsymbol{x}^{(b)},t_m) = \frac{|\Omega|}{I}\sum_{i=1}^{I}\sum_{n=1}^N \frac{1}{N} \boldsymbol{K}(\boldsymbol{x}^{(b)}-\boldsymbol{X}^n_{t_m}(\boldsymbol{\xi}^{(i)}))\omega(\boldsymbol{\xi}^{(i)},0)$\;
    $\mathcal{L} = \mathcal{L} + \sum_{b=1}^{B} \|\boldsymbol{u}_{\text{NN}}(\boldsymbol{x}^{(b)}, t_m)- \boldsymbol{\hat{u}}_{\text{NN}}(\boldsymbol{x}^{(b)},t_m)\|_2^2$\;
}
Update $\boldsymbol{u}_{\text{NN}}$'s parameters: $\boldsymbol{\Theta}_{t_m} =  \text{optim.Adam}(\boldsymbol{\Theta}_{t_m}, \nabla_{\boldsymbol{\Theta}_{t_m}} \mathcal{L});$ for $m=1,\cdots,M.$
}
\end{algorithm}

\subsubsection{Inverse problem}
Given the initial vortex $\omega_0$ and the dataset $\mathcal{D}: \{\boldsymbol{x}^{(d)},t^{(d)}; \boldsymbol{u}^{(d)}\}_{d=1}^D$ generated from a system that satisfies an NSE, the aim of the inverse problem is to infer unknown parameters from the data. A na\"ive approach is to add the data term directly to the loss function [Eq.~(\ref{loss})], which is consistent with the methodology in PINN. \cite{raissi2019physics} However, this approach has two drawbacks. On the one hand, we need to tune the hyperparameter carefully, as it balances the equation term and the data term. If the data are badly corrupted, the data term may unduly influence the equation term so that the model fails to learn the equation. On the other hand, we have to retrain a neural network again if we need to infer parameters from other data, which is time-consuming.

To address these problems, instead of directly adding the data term to the loss function, we devised a novel inference regime based on the following two procedures. First, we train a parametric solver network $\boldsymbol{u}_{\text{NN}}(\boldsymbol{x},\phi,t)$ that can generalize to different $\phi \in \Phi$, where $\phi$ is the unknown parameter in the parameter space $\Phi$. Second, we convert the inverse problem to the following optimization problem via the pretrained learning network for the parametric solver:
\begin{equation}\label{inv_loss}
\phi^* = \arg \min_{\phi \in \Phi} \sum_{d=1}^{D}\|\boldsymbol{u}_{\text{NN}}(\boldsymbol{x}^{(d)},\phi,t^{(d)})-\boldsymbol{u}^{(d)}\|_2^2.
\end{equation}

On the one hand, there is only one term in the loss function [Eq.~(\ref{inv_loss})]. Thus, we do not need to tune any hyperparameters in the loss function. In addition, due to the pretrained parametric solver network, we do not need to worry about information in the NSEs being corrupted by noise in the dataset. On the other hand, equipped with our pretrained parametric solver network, we need to optimize only the above low-dimension optimization problem in Eq.~(\ref{inv_loss}), which we do via the gradient descent algorithm. Thus, each inverse problem can be solved in seconds.

\section{Experiments}\label{sec4}
In this section, we apply DRVN to solve forward and inverse problems for incompressible NSEs. In this paper, all neural networks were initialized via the Xavier method. \cite{glorot2010understanding}  We used the relative $\ell_2$ error to evaluate the difference between the ground truth $\boldsymbol{u}$ and its prediction ${\boldsymbol{u}}_{\text{NN}}$. It is defined as $\|{\boldsymbol{x}}_{\text{NN}} - \boldsymbol{u}\|_2 / \|\boldsymbol{u}\|_2$. We evaluated the performance of our method with the relative $\ell_2$ error at the terminal time and with the mean relative $\ell_2$ error over the whole time interval, which we denote as ${E}_{T}$ and ${E}_{[0,T]}$, respectively. For each setting, we repeated the experiment five times with five different random number seeds and report the mean value and variance. We adopted Pytorch \cite{paszke2019pytorch} and TensorFlow \cite{abadi2016tensorflow} to implement DRVN and PINN, respectively. All experiments were implemented on an Nvidia GeForce RTX 3080Ti 12G, Nvidia GeForce RTX 3090 24G, and Nvidia Tesla V100 16G. All run times reported in this paper were evaluated on a GeForce RTX 3080Ti 12G.  When applying our method, we used a fully connected network with six hidden layers with equal hidden dimensions of 512 and used ReLU as the activation function. Tables~\ref{tab:detail_drvn} and~\ref{tab:detail_pinn} in Appendix~\ref{appendix:exp_details} contain further details of the setup of the experiments.

\subsection{Equations}
We conducted experiments on a Lamb--Oseen vortex, \cite{oseen1911wirbelbewegung} the fractional NSE, and a periodic Taylor--Green vortex. \cite{chorin1973numerical} Furthermore, we utilized DRVN to simulate a 3D Lamb--Oseen vortex, as described in Appendix~\ref{3d-lamb-oseen}. Table~\ref{setting} gives the kernel $\boldsymbol{K}(\boldsymbol{x})$ and driven noise in the corresponding SDE for NSEs with different dimensions and boundary conditions. More details of these equations will be introduced in the following sections.

\begin{table}[!ht]
\caption{Three NSEs studied in this paper.}\label{setting}
\centering
\begin{tabular}{cccc}
\hline
System & Domain & Kernel function $\boldsymbol{K}$ & Driven noise \\ 
\hline
Lamb--Oseen \cite{oseen1911wirbelbewegung}& \begin{tabular}[c]{@{}c@{}} $\mathbb{R}^2 $   \end{tabular} &  $\displaystyle \frac{1}{2\pi}\frac{\boldsymbol{x}^{\perp}}{\|\boldsymbol{x}\|_2^2}$   & Brownian
motion  \\[2ex]
     Fractional \cite{herrmann2011fractional}   & $\mathbb{R}^2$  & $\displaystyle \frac{1}{2\pi}\frac{\boldsymbol{x}^{\perp}}{\|\boldsymbol{x}\|_2^2}$ & L\'evy process \\[2ex]
Taylor--Green \cite{taylor1922diffusion} &       $[0,2\pi]^2$  & $\displaystyle \frac{1}{{4\pi}^2}\sum_{\boldsymbol{k}\in \mathbb{Z}^2} \frac{\boldsymbol{k}^{\perp}}{\|\boldsymbol{k}\|_2^2}\sin(\langle \boldsymbol{k}, \boldsymbol{x} \rangle)$   & Brownian motion \\\hline
\end{tabular}
\end{table}

\subsubsection{Lamb--Oseen vortex}
Here, we introduce the Lamb--Oseen vortex. Consider the following 2D vorticity equation:
\begin{equation}\label{vorticity_eq}
    \frac{\partial \omega(\boldsymbol{x}, t)}{\partial t}+\boldsymbol{u}(\boldsymbol{x}, t) \cdot \nabla \omega(\boldsymbol{x}, t)=\nu \Delta \omega(\boldsymbol{x}, t), 
\end{equation}
where the velocity field $\boldsymbol{u}(\boldsymbol{x},t)$ is given by the Biot--Savart law:
\begin{equation}
    \boldsymbol{u}(\boldsymbol{x}, t)=\frac{1}{2 \pi} \int_{\mathbb{R}^{2}} \frac{(\boldsymbol{x}-\boldsymbol{\xi})^{\perp}}{\|\boldsymbol{x}-\boldsymbol{\xi}\|_2^{2}} \omega(\boldsymbol{\xi}, t) \mathrm{d} \boldsymbol{\xi}.
\end{equation}
When the initial vorticity $\omega(\boldsymbol{x},0)=\alpha \delta(\boldsymbol{x})$, where $\delta(\boldsymbol{x})$ is Dirac's delta function, we can obtain the unique analytical solution of Eq.~(\ref{vorticity_eq}) as follows:
\begin{equation}
\omega(\boldsymbol{x}, t)=\frac{\alpha}{\nu t} G\left(\frac{\boldsymbol{x}}{\sqrt{\nu t}}\right), \quad \boldsymbol{u}(\boldsymbol{x}, t)=\frac{\alpha}{\sqrt{\nu t}} \boldsymbol{v}^{G}\left(\frac{\boldsymbol{x}}{\sqrt{\nu t}}\right),
\end{equation}
where the vorticity and velocity profiles are given by:
\begin{equation}
G(\boldsymbol{\xi})=\frac{1}{4 \pi} e^{-\|\boldsymbol{\xi}\|_2^{2} / 4}, \quad \boldsymbol{v}^{G}(\boldsymbol{\xi})=\frac{1}{2 \pi} \frac{\boldsymbol{\xi}^{\perp}}{\|\boldsymbol{\xi}\|_2^{2}}\left(1-e^{-\|\boldsymbol{\xi}\|_2^{2} / 4}\right). 
\end{equation}

We considered a computational domain of $[-2, 2]\times[-2, 2]$ and a time horizon of $[0, 1]$. The time step was $1/40$~s, and the Reynolds number was fixed as 10. 

\subsubsection{Fractional NSE}
We consider the 2D fractional NSEs [Eq.~(\ref{frac_omega})] described in Section~\ref{sec2.2}. Compared with the diffusion process in the RVM for the general 2D NSE, we just need to replace the Brownian motion with the L\'evy process for the fractional equation. The corresponding diffusion process and probabilistic representation are given by: \cite{zhang2012stochastic,zhang2012sfde}
\begin{equation}
    \begin{aligned}
    d\boldsymbol{X}_t&=\boldsymbol{u}(\boldsymbol{X}_t,t)dt+(2\nu)^{\frac{1}{\alpha}}d\boldsymbol{L}_t^{\alpha},\\
    \boldsymbol{u}(\boldsymbol{x},t)&=\int_{\mathbb{R}^2}\mathbb{E}[\boldsymbol{K}(\boldsymbol{x}-\boldsymbol{X}_t(\boldsymbol{\xi}))]\omega(\boldsymbol{\xi},0)d\boldsymbol{\xi}.
\end{aligned}
\end{equation}
where $\boldsymbol{L}_t^{\alpha}$ is the 2D $\alpha$-stable L\'evy process, and 
\[
\boldsymbol{K}(\boldsymbol{x}):=\frac{1}{2\pi}\frac{\boldsymbol{x}^{\perp}}{\|\boldsymbol{x}\|_2^{2}}.
\]

We considered a flow with a non-smooth initial condition given by
\begin{align*}
(\boldsymbol{\xi}^{(1)}, \omega(\boldsymbol{\xi}^{(1)},0)) &= ((0.5, 0.5), -0.2),   \\
(\boldsymbol{\xi}^{(2)}, \omega(\boldsymbol{\xi}^{(2)},0)) &= ((-0.5, 0.5), -0.2),  \\
(\boldsymbol{\xi}^{(3)}, \omega(\boldsymbol{\xi}^{(3)},0)) &= ((-0.5, -0.5), -0.2), \\
(\boldsymbol{\xi}^{(4)}, \omega(\boldsymbol{\xi}^{(4)},0)) &= ((0.5, -0.5), -0.2),  \\ 
(\boldsymbol{\xi}^{(5)}, \omega(\boldsymbol{\xi}^{(5)},0)) &= ((0.0, 0.0), 0.2). 
\end{align*}
The Reynolds number was fixed as 10.  The computational domain was $[-2, 2]\times[-2, 2]$ and the time horizon $[0, 1]$. To evaluate the performance of our method, we utilized the fine-grained RVM (Appendix~\ref{appendix:rvm}) as the ground truth. We divided the time period into 200 intervals and calculated the average of 1000000 independent paths generated by L\'evy processes. We utilized CUDA to accelerate the RVM algorithm. It took 4.5 h for an Nvidia GeForce RTX 3080Ti to generate each ground truth. For forward problems, we uniformly divided the total time period into 40 time intervals, i.e., $M=40$.

\subsubsection{Taylor--Green vortex}
A Taylor–Green vortex \cite{chorin1973numerical} is an exact solution to a 2D incompressible NSE with a periodic boundary condition in the domain $(\boldsymbol{x}_1, \boldsymbol{x}_2)\in [0, 2\pi]^2$, where its velocity $(u,v)$ and vorticity $\omega$ are given by:
\begin{equation}
\begin{aligned}
   {u}(\boldsymbol{x},t) &= \cos \boldsymbol{x}_1 \sin \boldsymbol{x}_2 e^{-2 \nu t},\\
    v(\boldsymbol{x},t) &= -\sin \boldsymbol{x}_1 \cos \boldsymbol{x}_2 e^{-2 \nu t},\\
    \omega(\boldsymbol{x},t) &= -2 \cos \boldsymbol{x}_1 \cos \boldsymbol{x}_2 e^{-2 \nu t},
\end{aligned}
\end{equation}
and the kernel functional of the probabilistic representation for a periodic equation in $[0, 2\pi]^2$ is given by:
\begin{equation}\label{periodic_kernel}
    K(\boldsymbol{x}) =\frac{1}{{4\pi}^2}\sum_{\boldsymbol{k}\in \mathbb{Z}^2/\{0\}} \frac{\boldsymbol{k}^{\perp}}{\|\boldsymbol{k}\|_2^2}\sin(\langle \boldsymbol{k}, \boldsymbol{x} \rangle).
\end{equation}

We divided the domain $\Omega$ into $64\times64$ grid elements and the time interval into 100 uniform intervals. The Reynolds number was 1. Notice that the kernel function in Eq.~(\ref{periodic_kernel}) is represented by an infinite series. Thus, we truncated Eq.~(\ref{periodic_kernel}) at $k_{\max}=10$.

\subsection{Baselines}\label{baseline}
For forward problems, we used PINNs as the baseline methods. PINNs optimize the FNN via the loss function, which includes the residual of the PDEs and the residual of the initial and boundary conditions. Given the initial data at $t_0$, i.e., $\{\omega(\boldsymbol{x}^{(i)},0), \boldsymbol{u}(\boldsymbol{x}^{(i)},0)\}_{i=1}^{N_1}$ with $\boldsymbol{x}^{(i)}\in \Omega$, and the boundary data  $\{\omega(\widehat{\boldsymbol{x}}^{(j)},t^{(j)}), \boldsymbol{u}(\widehat{\boldsymbol{x}}^{(j)},t^{(j)})\}_{j=1}^{N_2}$ with $\widehat{\boldsymbol{x}}^{(j)}\in\partial \Omega$, the loss function of the PINN in the vortex--velocity formulation is
\begin{multline}
\mathcal{L}_{\text{PINN}} = \sum_{i=1}^{N_1}\left(|\omega_{\text{NN}}(\boldsymbol{x}^{(i)},0)-\omega(\boldsymbol{x}^{(i)},0)|^2+\|\boldsymbol{u}_{\text{NN}}(\boldsymbol{x}^{(i)},0)-\boldsymbol{u}(\boldsymbol{x}_i,0)\|_2^2\right)+\\
\lambda_1\left(\left|\frac{\partial\omega_{\text{NN}}}{\partial t}+(\boldsymbol{u}_{\text{NN}}\cdot\nabla)\omega-\nu\Delta\omega_{\text{NN}}\right|^2+\left|\omega_{\text{NN}}-\nabla\times\boldsymbol{u}_{\text{NN}}\right|^2\right)+\\
\lambda_2\sum_{j=1}^{N_2}\left(|\omega_{\text{NN}}( \widehat{\boldsymbol{x}}^{(j)},t^{(j)})-\omega( \widehat{\boldsymbol{x}}^{(j)},t^{(j)})|^2+
\|\boldsymbol{u}_{\text{NN}}( \widehat{\boldsymbol{x}}^{(j)},t^{(j)})-\boldsymbol{u}( \widehat{\boldsymbol{x}}^{(j)},t^{(j)})\|_2^2\right),
\label{PINN1}
\end{multline}
where the input of the neural network is $(\boldsymbol{x},t)$ and the output is $(\boldsymbol{u}_{\text{NN}},\omega_{\text{NN}})$. Furthermore, PINNs represent the velocity field as the partial derivative of some latent function to satisfy automatically the divergence-free condition in the NSEs. \cite{raissi2019physics, Hendriks2020LinearlyCN}

In this paper, we consider two types of loss function, namely PINN and PINN+. The first has no access to the boundary data (PINN), i.e., $\lambda_2=0$ in Eq.~(\ref{PINN1}), whereas the second can utilize additional boundary data to encode the geometrical information (PINN+). For a periodic PDE when no additional boundary data are provided, we embed the periodic information in the PINN as follows. First, we constrain the left (lower) boundary to be equal to the right (upper) one, i.e., we replace the third term in the loss function [Eq.~(\ref{PINN1})] with the following boundary loss:
\begin{multline}
    \mathcal{L}_{\text{PINN}}^{bound} = \lambda_2\sum_{j=1}^{N_2}\left(\right.|\omega_{\text{NN}}( \widehat{\boldsymbol{x}}_{l}^{(j)},t^{(j)})-\omega_{\text{NN}}( \widehat{\boldsymbol{x}}_{r}^{(j)},t^{(j)})|^2
    +|\omega_{\text{NN}}( \widehat{\boldsymbol{x}}_{low}^{(j)},t^{(j)})-\omega_{\text{NN}}( \widehat{\boldsymbol{x}}_{up}^{(j)},t^{(j)})|^2\\
  +  \|\boldsymbol{u}_{\text{NN}}( \widehat{\boldsymbol{x}}_l^{(j)},t^{(j)})-\boldsymbol{u}_{\text{NN}}( \widehat{\boldsymbol{x}}_r^{(j)},t^{(j)})\|_2^2
  +   \|\boldsymbol{u}_{\text{NN}}( \widehat{\boldsymbol{x}}_{low}^{(j)},t^{(j)})-\boldsymbol{u}_{\text{NN}}( \widehat{\boldsymbol{x}}_{up}^{(j)},t^{(j)})\|_2^2\left.\right),   
\label{boundary_loss}
\end{multline}
where $\widehat{\boldsymbol{x}}_l$, $\widehat{\boldsymbol{x}}_r$, $\widehat{\boldsymbol{x}}_{low}$, and $\widehat{\boldsymbol{x}}_{up}$ denote data points sampled from the left, right, lower, and upper boundaries of $\Omega$, respectively. 

For inverse problems, we used the adjoint random vortex method (ARVM) as the baseline. Since RVM is a kind of differentiable solver, we can utilize the adjoint method \cite{pontryagin1987mathematical} to solve the inverse problem. Appendix~\ref{appendix:rvm} gives further details and the experimental setting. 

\subsection{Results: Forward problems}
In this section, we describe the results of experiments to solve three 2D forward problems: a Lamb--Oseen vortex, the fractional NSE, and a Taylor--Green vortex. Furthermore, we simulated a 3D Lamb--Oseen vortex, as described in Appendix~\ref{3d-lamb-oseen}.

\subsubsection{Lamb--Oseen Vortex}\label{forward:2-d lamb-oseen}
\paragraph{Comparison with PINNs.}
First, we simulated a 2D Lamb--Oseen vortex using our proposed method and the two PINNs. For the PINNs, we adopted the network architecture in NSFnets, \cite{jin2021nsfnets} which has seven hidden layers with 100 neurons and $\tanh$ activation functions. We looked at the two types of PINN introduced in Section~\ref{baseline}, namely PINN and PINN+. Moreover, as the initial condition approaches infinity around the origin, the sampling points for a PINN started from $t=0.025$~s to avoid extremely large values.

The relative $\ell_2$ errors of DRVN and the PINNs are given in Table~\ref{tab:2d_forward}. We can see that the DRVN simulation was accurate with an average relative $\ell_2$ error of $0.43\%$ for $t=T$ and $0.35\%$ for $t\in[0,T]$. Furthermore, both PINN and PINN+ performed poorly for the 2D Lamb--Oseen vortex. There are two main reasons why the PINNs failed. First, they failed to learn this ill-conditioned equation due to the singularity at $t=0$, which was also observed in \cite{krishnapriyan2021characterizing}. Second, the PINNs cannot embed the initial conditions on $\mathbb{R}^2$ integrally and receive only truncated initial information.

\begin{table}[!ht]
\caption{Comparison of relative $\ell_2$ error between DRVN and the PINNs for a 2D Lamb--Oseen vortex via deep RVM.}\label{tab:2d_forward}
\centering
\begin{tabular}{cccc}
\hline
Error      & DRVN & PINN & PINN+ \\ 
\hline
${E}_{T}\% $     & $0.43 \pm 0.01$ & $41.21 \pm 17.50$ & $12.24 \pm 4.18$ \\
${E}_{[0,T]}\% $ & $0.35 \pm 0.01$ & $47.39 \pm 13.81$ & $24.98 \pm 1.66$ \\\hline
\end{tabular} 
\end{table}

\begin{figure}[!htbp]
\begin{center}
\centerline{\includegraphics[width=1.0\linewidth]{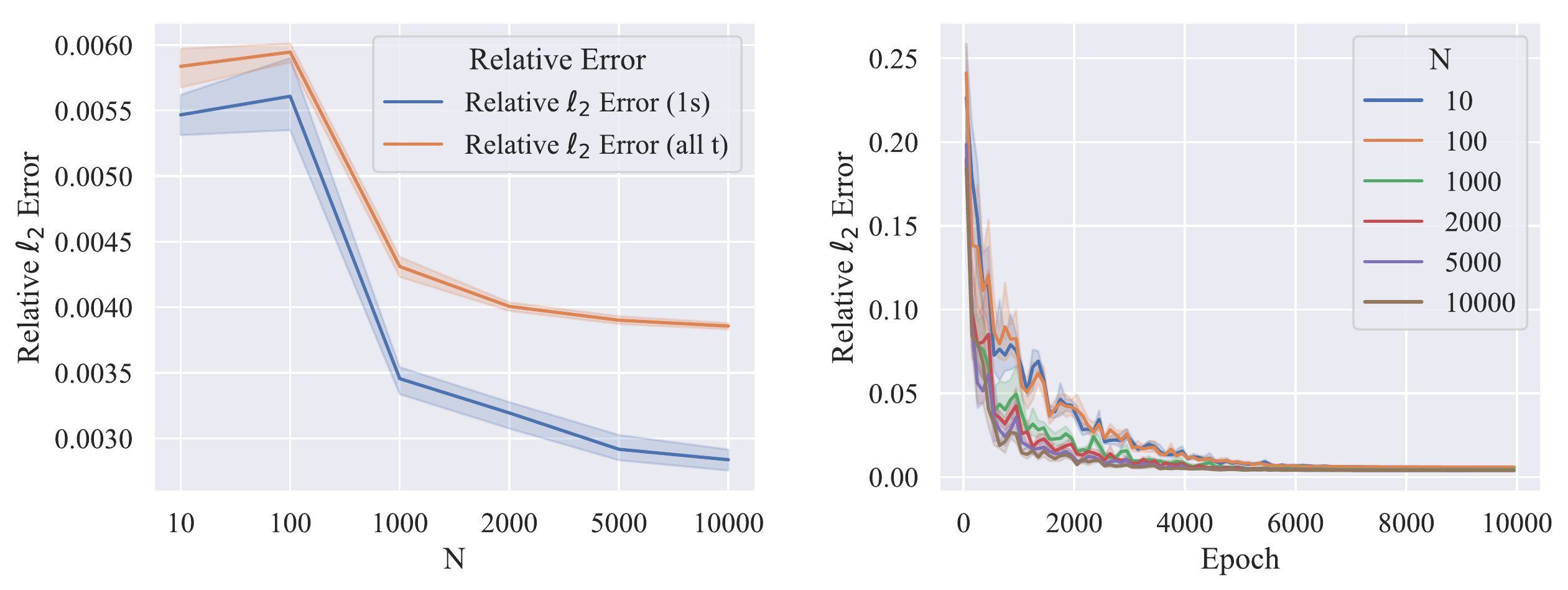}}
\caption{Results for a 2D Lamb--Oseen vortex for different numbers of samples. Left: Comparison of the relative $\ell_2$ error. Right: Comparison of the convergence of the relative $\ell_2$ error.}\label{lamb_oseen_error_N}
\end{center}
\end{figure}

\paragraph{Effects of $N$.} To test how changing $N$ affects the performance of our model, we evaluated our method with values of $N$ ranging from $10$ to ${10000}$. The relative $\ell_2$ errors and training dynamics are given in Fig.~\ref{lamb_oseen_error_N}. As we can see, with an increase in the number of sampling points, the error decreased, but the overall fluctuation of the error for $N>1000$ was smaller than $0.1\%$, which shows the robustness of DRVN over $N$. 

\begin{figure}[!htbp]
\begin{center}
\centerline{\includegraphics[width=1.0\linewidth]{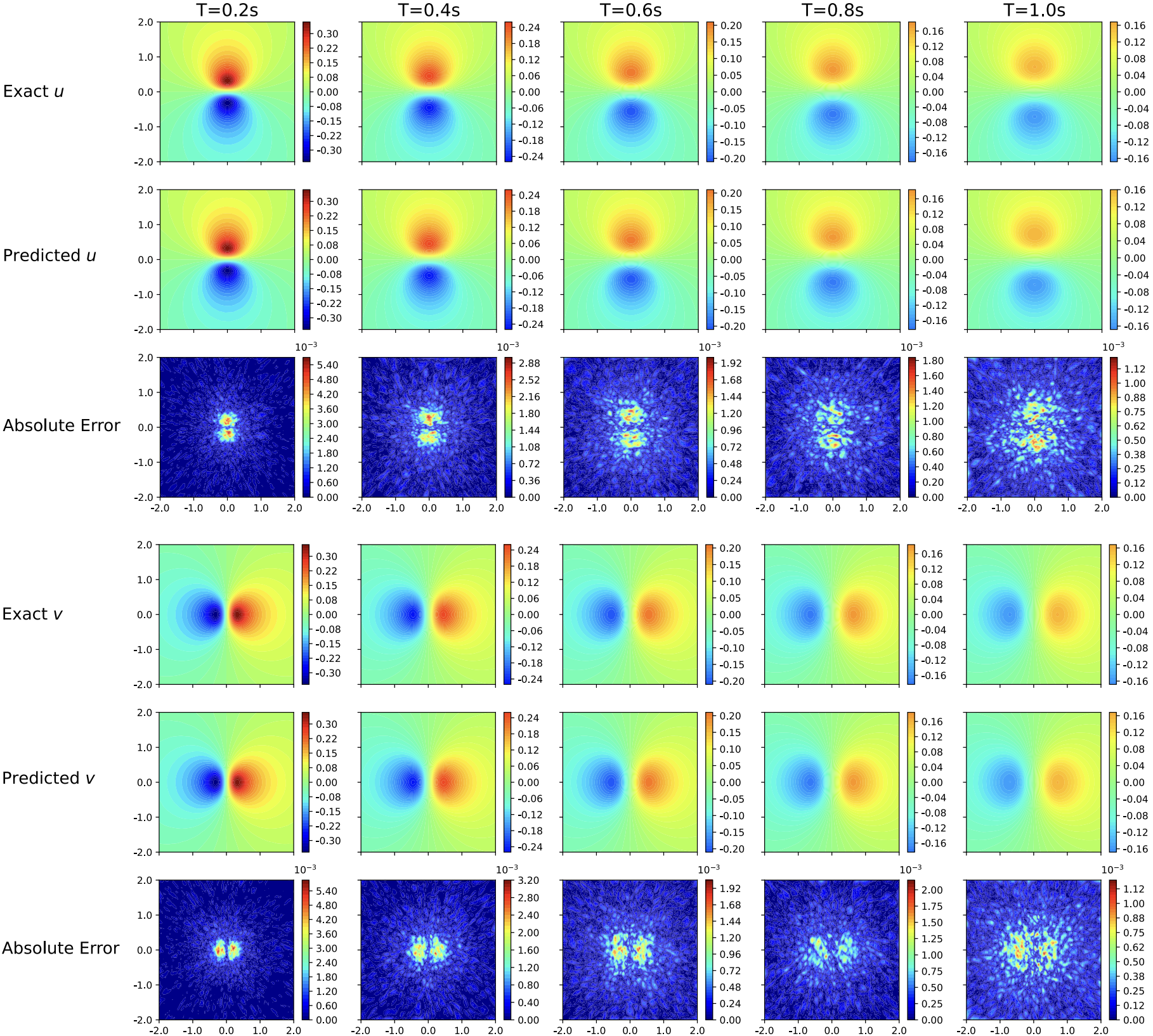}}
\caption{Comparison of velocity fields between the exact solution and deep RVM for a 2D Lamb--Oseen vortex from $T=0.2$ to $1.0$~s in $[-2,2]^2$.}\label{2d_lamb_oseen_forward}
\end{center}
\end{figure}

\begin{figure}[!htbp]
\begin{center}
\centerline{\includegraphics[width=1.0\linewidth]{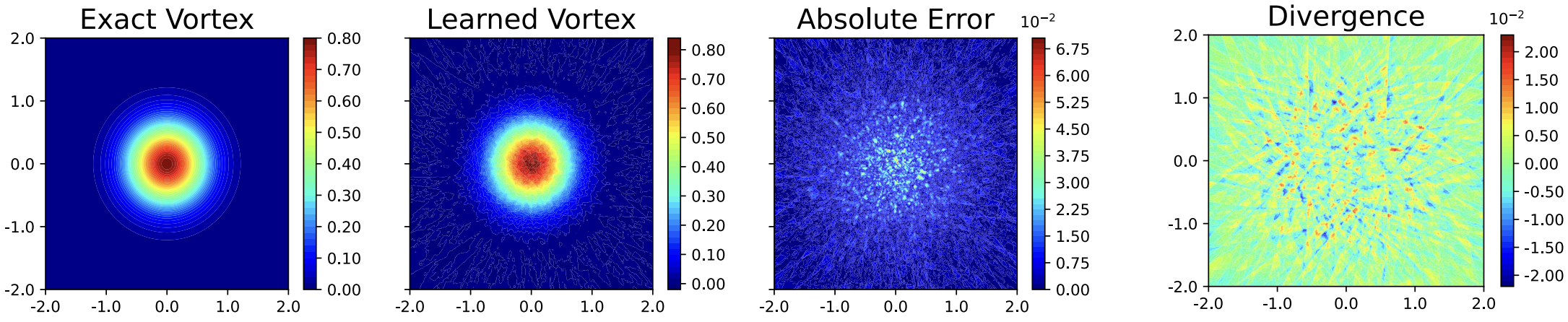}}
\caption{Results for a 2D Lamb--Oseen vortex. Left: Comparison of vorticity fields between the exact solution and the curl of the neural network at $T=1.0$~s in $[-2,2]^2$ (relative $\ell_2$ error $=$ 4.15\%). Right: Divergence of the neural network at $T=1.0$~s in $[-2,2]^2$ (mean absolute error $=$ 0.0031).}\label{curl}
\end{center}
\end{figure}

\paragraph{Meaningful derivative.} Snapshots of the learned velocity fields and corresponding absolute error for $T \in [0,1]$ are displayed in Fig.~\ref{2d_lamb_oseen_forward}. Note that we did not constrain the curl and divergence of $\boldsymbol{u}_{\text{NN}}$ explicitly in either Eq.~(\ref{nse_v}) or~(\ref{nse_omega}), but the neural network surprisingly learned the meaningful curl and divergence (Fig.~\ref{curl}), which correspond to the incompressible property and  vorticity--velocity formulation in the NSEs, respectively. This experimental result indicates that the neural network learned via DRVN has good physical properties.

\paragraph{Parametric solver learning.} Besides learning one model for a specific $\nu$, we also learned a parametric solver $\boldsymbol{u}(\boldsymbol{x},\nu,t)$, which can generalize to different values of $\nu$ ranging from $0.01$ to $0.5$. We changed the input dimension to 3 in the learning network for the parametric solver, and set the range of $\nu$ as $[0.001, 0.6]$ to enhance the generalization at the boundary. To distinguish between different values of $\nu$ for small orders of magnitude, we fed $\log{\nu}$ to the neural network rather than $\nu$ directly when training the parametric solver. The evaluation results are listed in Table~\ref{tab:operator40}. The relative $\ell_2$ errors are robust over $\nu$. The errors slightly increased as $\nu$ decreased, which is reasonable because the velocity field became sharper as $\nu$ increased.

\begin{table}[!ht]
\centering
\caption{Comparison of relative $\ell_2$ error for different values of $\nu$ in a parametric solver learning for a 2D Lamb--Oseen vortex.} \label{tab:operator40}
\begin{tabular}{lrr}
\hline
$\nu$ & ${E}_{T}$ (\%)   & ${E}_{[0,T]}$ (\%) \\
\hline
0.01  & $1.53 \pm 0.08$ & $4.00 \pm 0.14$ \\
0.02  & $1.41 \pm 0.07$ & $1.94 \pm 0.03$ \\
0.05  & $1.07 \pm 0.04$ & $1.35 \pm 0.01$ \\
0.1   & $0.91 \pm 0.03$ & $1.06 \pm 0.003$ \\
0.2   & $0.80 \pm 0.02$ & $0.91 \pm 0.01$ \\
0.5   & $0.87 \pm 0.05$ & $0.92 \pm 0.001$ \\
\hline
\end{tabular}
\end{table}

\paragraph{Training trick: Gradient stopping.} When training DRVN, we adopted gradient stopping as a training trick. All the neural networks $\{\boldsymbol{u}_{\text{NN}}(\boldsymbol{x},t_j)\}_{j=1}^{m-1}$ participate in calculating $X_{t_m}(\boldsymbol{\xi}^{(i)})$ in Eq.~(\ref{eluer}). Thus, the back propagation can become extremely time and memory consuming due to the composition of several neural network functions. To make the training more efficient and stable, we stopped the gradient flows in the computational graphs when updating $\boldsymbol{X}_{t_m}(\boldsymbol{\xi}^{(i)})$ during training, which is like what is done in reinforcement learning \cite{10.5555/3016100.3016191} and contrastive learning. \cite{chen2021exploring}

We ran ablation experiments to evaluate the effects of this gradient stopping trick. Table~\ref{gs} compares the relative error, required time, and memory with and without gradient stopping. We set the number of samples to 1000. The other settings were the same as for the forward problem described in Section~\ref{sec3.1}.  The gradient stopping trick reduces the time by around 37\% and the memory by around 36\% with comparable precision compared with the original training technique.

\begin{table}[!ht]
\caption{Results of ablation experiments to evaluate the effects of gradient stopping for a 2D Lamb--Oseen vortex.}\label{gs}
\centering
\begin{tabular}{lcc}
\hline
Parameter        & With gradient stopping              & Without           \\
\hline
${E}_{[0,T]}$ (\%) & $0.43 \pm 0.01$ & $0.42 \pm 0.004$ \\
${E}_{T}$ (\%)     & $0.35 \pm 0.01$ & $0.31 \pm 0.006$ \\
Time per epoch (s) &        $0.093$  &         $0.148$  \\
Memory (MB)        &          $3811$ &         $5987$   \\
\hline
\end{tabular}
\end{table}

\subsubsection{Fractional NSE}
\paragraph{General forward problems.} The diffusion parameter $\alpha$ is an essential property of the fractional equation. Thus, we evaluated our simulation algorithm for values of $\alpha$ ranging from 0.5 to 2. Table~\ref{tab:frac_pde} shows that the relative $\ell_2$ errors increased as $\alpha$ decreased. The errors at $T=1$~s were smaller than those for the whole interval. 

\begin{table}[!htbp]
\caption{Comparison of the relative $\ell_2$ errors for different values of $\alpha$ for the 2D fractional NSE.}\label{tab:frac_pde}
\centering
\begin{tabular}{lrr}
\hline
$\alpha$ & ${E}_{T}$ (\%)     & ${E}_{[0,T]}$ (\%)  \\
\hline
0.50     & $13.02 \pm 1.05$ & $12.11 \pm 0.27$ \\
0.75     &  $5.97 \pm 0.43$ &  $7.84 \pm 0.28$ \\
1.00     &  $2.32 \pm 0.27$ &  $4.15 \pm 0.05$ \\
1.25     &  $1.15 \pm 0.18$ &  $3.00 \pm 0.05$ \\
1.50     &  $0.69 \pm 0.07$ &  $1.71 \pm 0.05$ \\
1.75     &  $0.81 \pm 0.09$ &  $1.10 \pm 0.01$ \\
2.00     &  $0.60 \pm 0.05$ &  $0.87 \pm 0.01$ \\\hline
\end{tabular}
\end{table}

\begin{figure}[!htbp]
\begin{center}
\centerline{\includegraphics[width=0.93\linewidth]{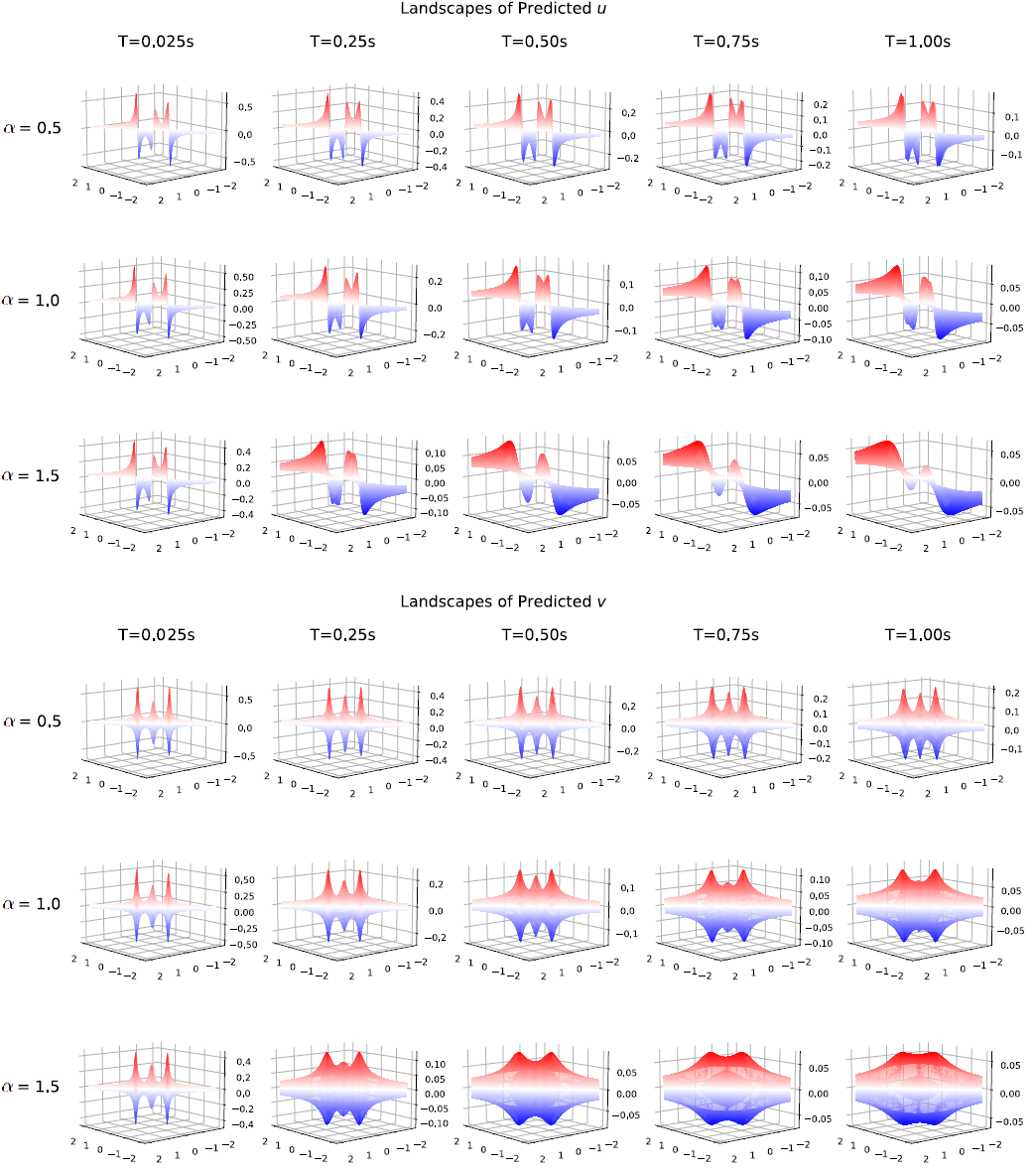}}
\caption{Comparison of learned landscapes for different values of $\alpha$ from $T=0.025$ to $1.00$~s in $[-2,2]^2$ for the 2D fractional NSE.}
\label{fig_landscape}
\end{center}
\end{figure}

\begin{figure}[!htbp]
\begin{center}
\centerline{\includegraphics[width=0.95\linewidth]{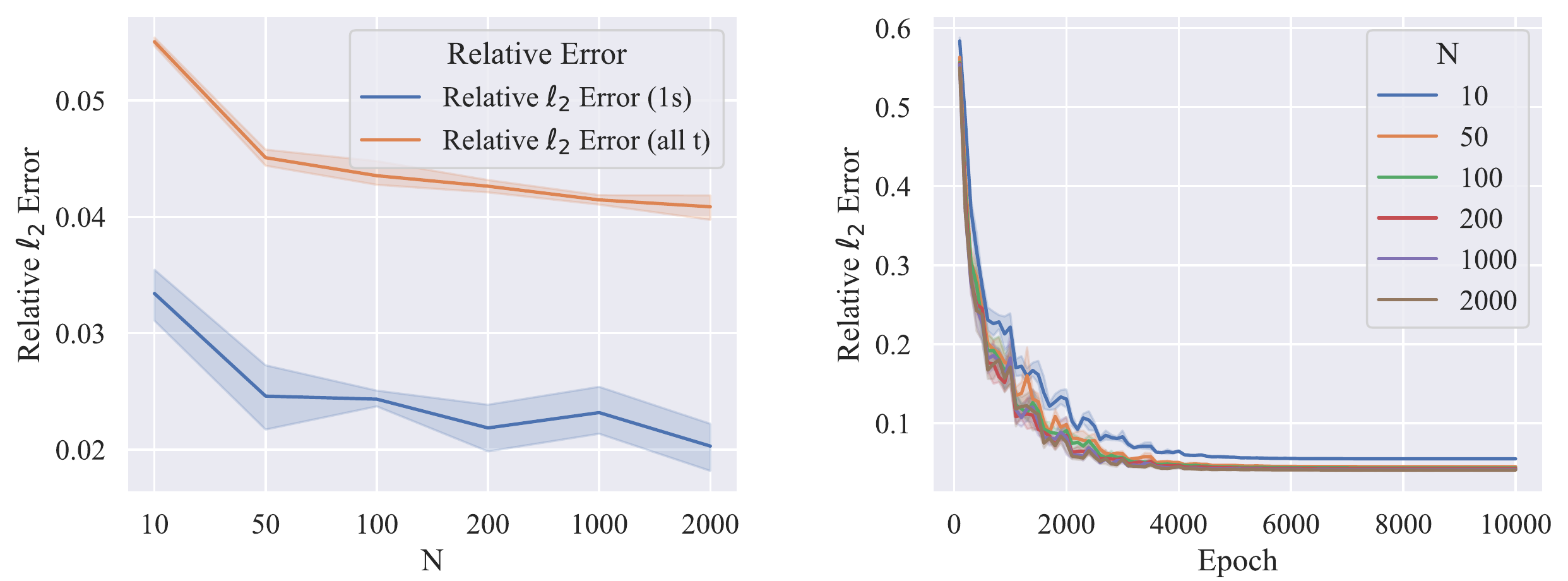}}
\caption{Results for the 2D fractional NSE for different numbers of samples. Left: Comparison of the relative $\ell_2$ error. Right: Comparison of the convergence of the relative $\ell_2$ error.}
\label{frac_error}
\end{center}
\end{figure}

To gain a deeper understanding of the influence of $\alpha$, we plot the landscapes of the learned fractional equations from $T=0.025$ to $1.00$~s in Fig.~\ref{fig_landscape}. On the one hand, the surfaces of the equations became more singular as $\alpha$ became smaller. Thus, training the neural networks became more difficult, which is consistent with the frequency principle of neural networks studied in \cite{xu2019frequency, pmlr-v97-rahaman19a}. On the other hand, the landscapes became smoother over time. Thus, the error at the terminal time was relatively smaller than over the whole interval. Furthermore, we show the effects of the number of sampling paths in Fig.~\ref{frac_error}. In general, the error decreased as we increased the number of samples~$N$.

\paragraph{Parametric solver learning.} We also learned a parametric solver $\boldsymbol{u}(\boldsymbol{x},\alpha,t)$, which can generalize to values of $\alpha$ ranging from $1.00$ to $2.00$. The results are shown in Table~\ref{tab:operator_frac}. The relative $\ell_2$ errors increased when $\alpha$ decreased, as they did for the forward problem.

\begin{table}[!htbp]
\caption{Comparison of the relative $\ell_2$ errors for different values of $\alpha$ for parametric solver learning for the 2D fractional NSE.}\label{tab:operator_frac}
\centering
\begin{tabular}{ccc}
\hline
$\alpha$ & ${E}_{T}$ (\%)  & ${E}_{[0,T]}$ (\%) \\
\hline
1.00     & $4.21 \pm 0.29$ & $7.09 \pm 0.07$    \\
1.25     & $1.90 \pm 0.04$ &    $3.69 \pm 0.08$ \\
1.50     & $1.67 \pm 0.06$ &    $2.66 \pm 0.04$ \\
1.75     & $1.41 \pm 0.05$ &    $2.99 \pm 0.05$ \\
2.00     & $1.35 \pm 0.08$ &    $2.54 \pm 0.10$ \\
\hline
\end{tabular}
\end{table}

\subsubsection{Taylor--Green vortex}
\paragraph{Comparisons with the PINNs.} In this part, we simulate a 2D Taylor--Green vortex using our proposed method and the two PINNs. Like the experiments in Section~\ref{forward:2-d lamb-oseen}, we use PINN and PINN+ as the baselines. We choose a fully connected $\tanh$ neural network with seven hidden layers and with 500 neurons per layer. \cite{jin2021nsfnets}  Table~\ref{tg_error_pinn} shows that PINN fails to simulate the velocity field because no boundary information is provided. Note that the results for DRVN are comparable with those for PINN+, although the loss function of DRVN does not include boundary data during $t\in(0,T]$ as a term.

\begin{table}[!htbp]
\caption{Comparison of the relative $\ell_2$ errors between DRVN and the PINNs for a 2D Taylor--Green vortex.}\label{tg_error_pinn}
\centering

\begin{tabular}{lcc}
\hline
Algorithm    & ${E}_{T}$ (\%)     & ${E}_{[0,T]}$ (\%) \\
\hline
DRVN  & $1.67 \pm 0.04$    & $0.94 \pm 0.03$    \\
PINN  & $102.08 \pm 47.12$ & $35.31 \pm 14.94$ \\
PINN+ & $1.78 \pm 0.40$    & $0.53 \pm 0.12$ \\\hline
\end{tabular}
\end{table}

\begin{figure}[!htbp]
\begin{center}
\centerline{\includegraphics[width=1.0\linewidth]{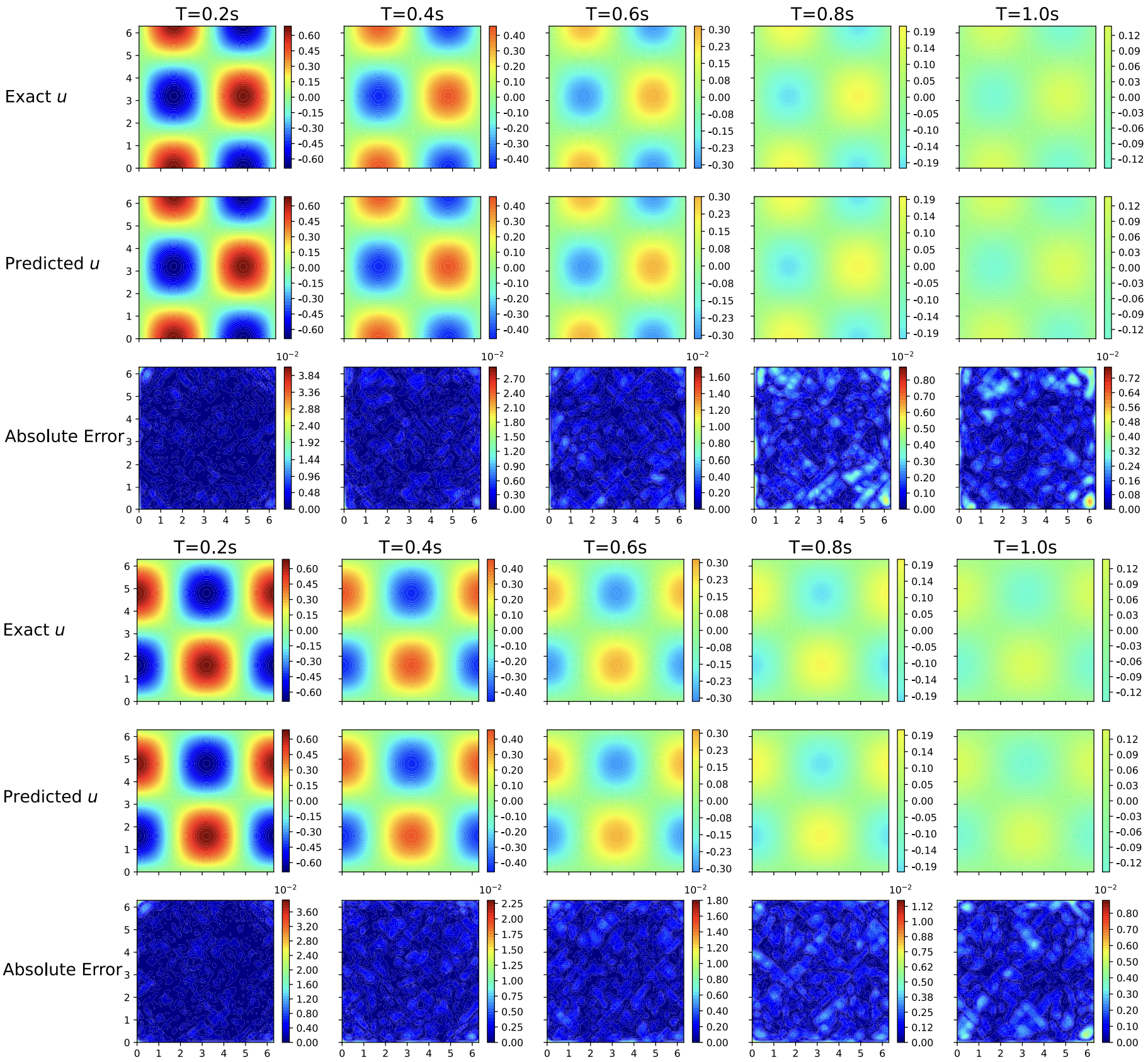}}
\caption{Comparison of velocity fields between the exact solution and deep RVM from $T=0.2$ to $1.0$~s in the periodic domain $[0,2\pi]^2$ for a 2D Taylor--Green vortex.}\label{u_tg}
\end{center}
\end{figure}

Snapshots of the learned velocity fields and corresponding absolute error during $T \in [0,1]$ are displayed in Fig.~\ref{u_tg}. As shown, DRVN can obtain accurate solutions for the Taylor--Green vortex. The errors are mainly at the domain boundary. 

\paragraph{Architecture trick: Periodic preprocessing operator.}
When solving periodic NSEs in a periodic domain $\Omega=[0,b_1]\times[0,b_2]$, we sample data points in $\Omega$ and then train the network. However, the paths of $\boldsymbol{X}_t$ in Eq.~(\ref{eluer}) can extend out of $\Omega$ due to the randomness of the Brownian motion, and the neural network was not trained out of $\Omega$. To reduce the resulting error, we utilized the periodic preprocessing operator to adjust the path of the SDE. We add the following periodic preprocessing operator $\boldsymbol{f}_{\text{pp}}$ before the forward propagation of the neural network:
\begin{equation}
    \boldsymbol{f}_{\text{pp}}(\boldsymbol{x}) = \left(\boldsymbol{x}_1 - \left\lfloor \frac{\boldsymbol{x}_1}{b_1}\right\rfloor b_1,\  \boldsymbol{x}_2 - \left\lfloor \frac{\boldsymbol{x}_2}{b_2}\right\rfloor b_2\right).
\end{equation}
Due to the periodicity of this equation, $\boldsymbol{u}( \boldsymbol{f}_{\text{pp}}(\boldsymbol{x}),t)=\boldsymbol{u}(\boldsymbol{x},t)$ for all $\boldsymbol{x}$ and $t$. Thus, the periodic preprocessing operator projects the SDE output inside the domain $\Omega$ without inducing additional errors.

The effects of this periodic preprocessing with different sizes of the temporal lattice $M$ are listed in Table~\ref{tg_error}. This periodic preprocessing trick improves the performance of DRVN significantly with no additional computational cost for all $M$.

\begin{table}[!htbp]
\caption{Comparison of relative $\ell_2$ errors for different numbers of time intervals $M$ in ablation experiments to evaluate the effects of periodic preprocessing (pp) for a 2D Taylor--Green vortex.}\label{tg_error}
\centering
\begin{tabular}{lccccc}
\hline
                       & $5$            & $10$           & $20$           & $40$           & $100$  \\
\hline
${E}_{T}$ (\%)         & ${2.43 \pm0.15}$ & ${1.90 \pm0.06}$ & ${1.71 \pm0.09}$ & ${1.71 \pm0.04}$ & ${1.67 \pm0.04}$ \\
${E}_{T}$ (\%) w/o pp     & $2.68 \pm0.11$ & $2.08 \pm0.15$ & $1.88 \pm0.05$ & $1.81 \pm0.06$ & $1.79 \pm0.06$ \\
${E}_{[0,T]}$ (\%)     & ${1.85 \pm0.05}$ & ${1.16 \pm0.07}$ & ${0.96 \pm0.04}$ & ${0.99 \pm0.10}$ & ${0.94 \pm0.03}$ \\
${E}_{[0,T]}$ (\%) w/o pp & $1.95 \pm0.01$ & $1.28 \pm0.09$ & $1.11 \pm0.04$ & $1.01 \pm0.04$ & $1.04 \pm0.06$ \\
Time (s)               &       $0.052$ &       $0.103$ &       $0.206$ &       $0.412$ &       $1.028$ \\
Time w/o pp (s)        &       $0.052$ &       $0.102$ &       $0.204$ &       $0.408$ &       $1.008$ \\\hline
\end{tabular}
\end{table}

\subsection{Results: Inverse problems}
In this section, we conduct inverse problems on a Lamb--Oseen vortex and the fractional NSE. We aim to infer the viscosity term $\nu$ and the diffusion parameter $\alpha$ from given data for these two equations, respectively. We consider the following three training datasets: 
\begin{enumerate}
    \item $A_1$ clean data points, $\{(\boldsymbol{x}^{(i)},t^{(i)}; \boldsymbol{u}(\boldsymbol{x}^{(i)},t^{(i)}))\}_{i=1}^{A_1}$
    \item $A_2$ noisy data points whose labels are perturbed by additive uncorrelated Gaussian noise, $\{(\boldsymbol{x}^{(i)},t^{(i)}; \boldsymbol{u}(\boldsymbol{x}^{(i)},t^{(i)})+\boldsymbol{\epsilon}^{(i)})\}_{i=1}^{A_2}$, where $\boldsymbol{\epsilon^{(i)}}\sim\mathcal{N}(\boldsymbol{0},(0.01\|\boldsymbol{u}\|_2)^2\cdot\boldsymbol{I}_{2})$
    \item $A_3$ noisy data points whose labels are perturbed by additive uncorrelated Gaussian noise, $\{(\boldsymbol{x}^{(i)},t^{(i)}; \boldsymbol{u}(\boldsymbol{x}^{(i)},t^{(i)})+\boldsymbol{\epsilon}^{(i)})\}_{i=1}^{A_3}$, where $\boldsymbol{\epsilon^{(i)}}\sim\mathcal{N}(\boldsymbol{0},(0.1\|\boldsymbol{u}\|_2)^2\cdot\boldsymbol{I}_{2})$
\end{enumerate}
All data points were sampled uniformly in $[-2,2]^2$ at $T=1$~s. We set $(A_1, A_2, A_3) = (100,100,1000)$ and $(100,1000,2000)$ for the Lamb--Oseen vortex and the fractional NSE, respectively.

\subsubsection{Lamb--Oseen vortex}\label{inver_lamb}
In this part, we apply the learned parametric solver network to handle inverse problems for the three training datasets. The results for DRVN and ARVM are reported in Table~\ref{tab:lamb_oseen_inverse} using subscripts $D$ and $A$, respectively. The table shows that, compared with ARVM, DRVN needs only 1\% of the time to solve the inverse problem, and obtains remarkably higher accuracy.

\begin{table}[!ht]
\caption{Comparison of the relative errors of inverse problems between DRVN and ARVM under various situations for a 2D Lamb--Oseen vortex for different values of $\nu$.  $\hat{\nu}_D $ ($\hat{\nu}_A$) and $E_D$ ($E_A$) represent the estimator and relative error of DRVN (ARVM), respectively. The final column is the total training time.}\label{tab:lamb_oseen_inverse}
\centering
\resizebox{\columnwidth}{!}{
\begin{tabular}{lrrrrrrr}
\hline
 & \multicolumn{1}{c}{1} & \multicolumn{1}{c}{2} & \multicolumn{1}{c}{5} & \multicolumn{1}{c}{10} & \multicolumn{1}{c}{20} & \multicolumn{1}{c}{50} & \multicolumn{1}{c}{Time (s)} \\
\hline
\multicolumn{8}{c}{100 clean data points} \\
$\hat{\nu}_{D}$                                                           &   $1.02 \pm0.06$ &   $2.00 \pm0.04$ & $5.00 \pm0.02$ &  $9.95 \pm0.02$ & $20.12 \pm0.05$ & $50.05 \pm0.07$ &   \multirow{2}{*}{$1.96$} \\
$E_D$ (\%)                                                                    &   $3.58 \pm4.46$ &   $1.51 \pm1.09$ & $0.30 \pm0.34$ &  $0.52 \pm0.19$ &  $0.61 \pm0.25$ &  $0.14 \pm0.09$ & \\[1ex]
$\hat{\nu}_A$                                                               &   $0.90 \pm0.42$ &   $1.15 \pm0.90$ & $5.03 \pm0.17$ & $10.08 \pm0.27$ & $20.60 \pm0.28$ & $52.40 \pm2.07$ & \multirow{2}{*}{$213.17$} \\
$E_A$ (\%)                                                                    & $35.58 \pm18.30$ & $42.56 \pm45.19$ & $2.47 \pm2.05$ &  $2.22 \pm1.44$ &  $2.99 \pm1.39$ &  $4.81 \pm4.14$ & \\
\\
\multicolumn{8}{c}{100 noisy data points with $1\%$ Gaussian noise} \\
$\hat{\nu}_D$                                                               &   $1.01 \pm0.03$ &  $2.01 \pm0.005$ & $4.97 \pm0.02$ &  $9.97 \pm0.05$ & $20.10 \pm0.03$ & $49.97 \pm0.08$ &   \multirow{2}{*}{$2.01$} \\
$E_D$ (\%)                                                                    &   $2.31 \pm2.01$ &   $1.80 \pm1.38$ & $0.55 \pm0.46$ &   $0.44 \pm034$ &  $0.50 \pm0.13$ &  $0.14 \pm0.09$ & \\[1ex]
$\hat{\nu}_A$                                                               &   $1.29 \pm0.90$ &   $1.87 \pm0.36$ & $5.01 \pm0.08$ & $10.04 \pm0.34$ & $20.43 \pm0.77$ & $50.97 \pm0.94$ & \multirow{2}{*}{$213.47$} \\
$E_A$ (\%)                                                                    & $67.75 \pm58.15$ &  $15.07 \pm9.77$ & $1.21 \pm0.84$ &  $2.64 \pm1.80$ &  $3.31 \pm2.66$ &  $1.93 \pm1.88$ & \\
\\
\multicolumn{8}{c}{1000 noisy data points with $10\%$ Gaussian noise} \\
$\hat{\nu}$                                                                 &   $1.01 \pm0.02$ &  $2.02 \pm0.021$ & $4.98 \pm0.04$ &  $9.99 \pm0.12$ & $20.14 \pm0.10$ & $50.10 \pm0.44$ &  \multirow{2}{*}{$2.36$} \\
$E_D$ (\%)                                                                    &   $1.82 \pm1.24$ &   $1.07 \pm0.74$ & $0.64 \pm0.49$ &  $0.87 \pm0.72$ &  $0.72 \pm0.49$ &  $0.67 \pm0.50$ & \\[1ex]
$\hat{\nu}_A$                                                               &   $0.66 \pm0.19$ &   $1.73 \pm0.08$ & $5.02 \pm0.06$ & $10.17 \pm0.08$ & $20.62 \pm0.22$ & $51.85 \pm0.60$ & \multirow{2}{*}{$215.36$} \\
$E_A$ (\%)                                                                    & $34.12 \pm18.99$ &  $13.72 \pm4.14$ & $0.95 \pm0.60$ &  $1.67 \pm0.79$ &  $3.09 \pm1.08$ &  $3.70 \pm1.21$ & \\\hline
\end{tabular}}
\end{table}

\subsubsection{Factional PDE}
Using the same settings described in Section~\ref{inver_lamb}, we apply the learned parametric solver network with respect to $\alpha$ to handle inverse problems for the three inverse problems mentioned above. The results are reported in Table~\ref{tab:frac_inverse}. The relative errors ranged from 0.001 to 0.01 under low noise levels, which are sufficiently small. Moreover, the relative errors were around 0.05 when the noise was high, indicating that more data were required to obtain a more precise estimate. Note that the total time for optimizing the inverse loss was less than 3~s for all situations.

\begin{table}[!htbp]
\caption{Comparison of the relative errors of inverse problems under various situations for the 2D fractional NSE for different values of the ground truth $\alpha$. $\hat{\alpha}$ and ${E}$ are the estimator and relative error, respectively. The final column is the total training time.}\label{tab:frac_inverse}
\centering
\begin{tabular}{ccccccc}
\hline
 & \multicolumn{1}{c}{1.00} & \multicolumn{1}{c}{1.25} & \multicolumn{1}{c}{1.50} & \multicolumn{1}{c}{1.75} & \multicolumn{1}{c}{2.00}         & Time (s) 
\\
\hline
\multicolumn{7}{c}{100 clean data points} \\
$\hat{\alpha}$                                                       &             $0.99 \pm0.02$ &            $1.26 \pm0.003$ &            $1.50 \pm0.004$ &            $1.75 \pm0.002$ & $1.99 \pm0.01$ & \multirow{2}{*}{$1.81$} \\
${E}$ (\%)                                                           &             $1.25 \pm0.81$ &             $0.73 \pm0.24$ &             $0.15 \pm0.23$ &             $0.07 \pm0.10$ &                     $0.49 \pm0.33$ &                          \\ 
\\
\multicolumn{7}{c}{1000 noisy data points with $1\%$ Gaussian noise} \\
$\hat{\alpha}$                                                       &             $0.99 \pm0.02$ &             $1.26 \pm0.01$ &             $1.49 \pm0.01$ &             $1.75 \pm0.01$ &                     $1.98 \pm0.03$ & \multirow{2}{*}{$2.07$} \\
${E}$ (\%)                                               &             $1.35 \pm1.09$ &             $0.57 \pm0.74$ &             $0.61 \pm 0.59$ & $0.65 \pm 0.39$ & $1.19 \pm 1.08$
 \\
\\
\multicolumn{7}{c}{2000 noisy data points with $10\%$ Gaussian noise} \\
$\hat{\alpha}$                                                       &             $1.03 \pm0.07$ &             $1.27 \pm0.08$ &             $1.49 \pm0.10$ &             $1.72 \pm0.10$ &                     $1.93 \pm0.12$ & \multirow{2}{*}{$2.61$} \\
${E}$ (\%)                                                           &             $4.87 \pm5.48$ &             $4.46 \pm4.81$ &             $5.85 \pm2.05$ &             $3.84 \pm4.75$ &                     $5.03 \pm3.98$ & \\\hline
\end{tabular}
\end{table}

\section{Conclusions}\label{sec5}
In this paper, we propose DRVN for simulating the fluids and inferring unknown parameters of NSEs. DRVN utilizes the probabilistic representation in the random vortex formulation of an NSE and substitutes Monte Carlo sampling for the derivative calculation. Thus, DRVN can solve non-smooth and fractional NSEs efficiently, which expands the application of the deep learning method in fluid mechanics. The numerical experiments on various equations verify our algorithm. However, DRVN still has some limitations. First, the convergence rate of DRVN is non-trivial due to the non-convexity of neural networks and the stopping gradient technique. Second, we do not consider NSEs with an external force, which is a critical situation in control. We will investigate both the convergence of DRVN and apply DRVN to NSEs with an external force in future work.

\section*{Acknowledgments}
This project was supported financially by Microsoft Research. We sincerely acknowledge Professor Xicheng Zhang from Wuhan University for helpful discussions on the RVM for the fractional NSE. We also sincerely acknowledge Professor Shihua Zhang from the Chinese Academy of Sciences for support and discussions.

\appendix

\section{Random vortex method}\label{appendix:rvm}
In this section, we take the 2D NSE as an example to illustrate RVM.  Recall the Euler discretization in Eq.~(\ref{X_dynamic}):
\begin{equation}\label{ap:eluer}
    \boldsymbol{X}_{t_{m}}(\boldsymbol{\xi}^{(i)}) - \boldsymbol{X}_{t_{m-1}}(\boldsymbol{\xi}^{(i)}) = \boldsymbol{u}( \boldsymbol{X}_{t_{m-1}}(\boldsymbol{\xi}^{(i)}), t_{m-1}) \Delta t + \sqrt{2 \nu} \Delta \boldsymbol{B}_{m}.
\end{equation}
RVM utilizes the probabilistic representation of the velocity field in Eq.~(\ref{pro_rep}) to calculate $\boldsymbol{u}( \boldsymbol{X}^{n^{\prime}}_{t}(\boldsymbol{\xi}^{(i)}), t)$:
\begin{equation}\label{ap:pro_rep}
\boldsymbol{u}( \boldsymbol{X}^{n^{\prime}}_{t}(\boldsymbol{\xi}^{(i)}), t) = \frac{|\Omega|}{I}\sum_{i=1}^{I}\sum_{n=1}^N \frac{1}{N} \boldsymbol{K}(\boldsymbol{X}^{n^{\prime}}_{t}(\boldsymbol{\xi}^{(i)})-\boldsymbol{X}^n_{t}(\boldsymbol{\xi}^{(i)}))\omega(\boldsymbol{\xi}^{(i)},0).
\end{equation}
For other points $\boldsymbol{x}$ out of the coordinate points $\{\boldsymbol{\xi}^{(i)}\}_{i=1}^{I}$, RVM calculates $\boldsymbol{u}(\boldsymbol{x}, t)$ as follows:
\begin{equation}\label{ap:crvm}
\boldsymbol{{u}}(\boldsymbol{x},t) = \frac{|\Omega|}{I}\sum_{i=1}^{I}\sum_{n=1}^N \frac{1}{N} \boldsymbol{K}(\boldsymbol{x}-\boldsymbol{X}^n_{t}(\boldsymbol{\xi}^{(i)}))\omega(\boldsymbol{\xi}^{(i)},0).
\end{equation}

Since RVM is a kind of differentiable solver, we can utilize the adjoint method for the inverse problem. Given the initial vortex $\omega_0$ and the dataset $\mathcal{D}: \{\boldsymbol{x}^{(d)},t^{(d)}, \boldsymbol{u}^{(d)}\}_{d=1}^D$ generated from a system that satisfies the NSEs, ARVM is based on the following optimization problem: 
\begin{equation}\label{ap:inv_loss}
\phi^* = \arg \min_{\phi \in \Phi} \sum_{d=1}^{D}\|\boldsymbol{u}_{\text{RVM}}(\boldsymbol{x}^{(d)},\phi,t^{(d)})-\boldsymbol{u}^{(d)}\|_2^2.
\end{equation}
As $\boldsymbol{u}_{\text{RVM}}$ is differentiable, we directly utilize Adam to find the optimum $\phi^*$. We optimize the loss function for 2000 epochs with the initial learning rate 0.01 and reduce the learning rate by a factor of 0.2 every 500 epochs. To distinguish between different values of $\nu$ with small orders of magnitude and obtain stable results, we feed $\log\sqrt{{\nu}}$ into ARVM rather than $\nu$ directly.

\section{Experimental Details}\label{appendix:exp_details}

\subsection{Experimental Details for DRVN}\label{appendix:detail_DRVN}

\begin{table}[H]
\caption{Experimental details for DRVN.}\label{tab:detail_drvn}
\centering
\begin{tabular}{lllrrrrr}
\hline
Equations                       & Problems   & $lr$   & Epoch & Step size & Decay rate & Batch size     & $N$  \\ 
\hline
\multirow{3}{*}{2D Lamb--Oseen} & Forward    & 0.001  & {10000} & 500       & 0.5        & 2000           & 1000 \\ 
                                & Parametric & 0.001  & {10000} & 500       & 0.5        & {20000}          & 500  \\  
                                & Inverse    & 0.01   & 2000  & 500       & 0.2        & Data size      & NA   \\ 
\\
\multirow{3}{*}{2D Fractional} & Forward    & 0.001  & {10000} & 500       & 0.5        & 2000           & 1000 \\ 
                                & Parametric & 0.001  & {10000} & 500       & 0.5        & {10000}          & 1000 \\
                                & Inverse    & 0.01   & 2000  & 500       & 0.2        & Data size      & NA   \\ 
\\
2D Taylor--Green                & Forward    & 0.0005 & {20000} & 1000      & 0.5        & 100            & 2    \\
3D Lamb--Oseen                  & Forward    & 0.0005 & {20000} & 1000      & 0.5        & 100            & 10   \\ \hline
\end{tabular}
\end{table}

\subsection{Experimental Details for PINNs}\label{appendix:detail_pinn}

\begin{table}[H]
\caption{Experimental details for PINNs.}\label{tab:detail_pinn}
\centering
\begin{tabular}{lllll}
\hline
            & \multicolumn{2}{c}{2D Lamb--Oseen} & \multicolumn{2}{c}{2D Taylor--Green} \\
Parameter   & PINN            & PINN+           & PINN            & PINN+           \\
\hline
$lr$        & 0.0001          & 0.0001          & 0.0001          & 0.0001          \\
Epoch       &  {10000}      &  {10000}      &  {20000}      & {20000}      \\
Step size   & 2000            & 2000            & {10000}      & {10000}           \\
Decay rate  & 0.1             & 0.1             & 0.1             & 0.1             \\
Batch size  & $2.6\times10^6$ & $2.6\times10^6$ & $8.1\times10^4$ & $8.1\times10^4$ \\
$N_1$       & $6.6\times10^4$ & $6.6\times10^4$ & $4.1\times10^3$ & $4.1\times10^3$ \\
$N_2$       & $4.1\times10^4$ & 0               & $5.1\times10^3$ & $5.1\times10^3$ \\
$\lambda_1$ & 100             & 100             & 1               & 1               \\
$\lambda_2$ & 0               & 100             & 1               & 1               \\ \hline
\end{tabular}
\end{table}

\section{3D Lamb--Oseen Vortex}\label{3d-lamb-oseen}
In this experiment, we aim to simulate the velocity field for a 3D Lamb--Oseen vortex, \cite{oseen1911wirbelbewegung} which is represented by a 3D incompressible NSE with initial vorticity $(0,0,\delta(\boldsymbol{x}_1)\delta(\boldsymbol{x}_2))$ for $\boldsymbol{x}\in\mathbb{R}^3$. The corresponding exact solution of the velocity field is given by:
\begin{equation}\label{3d_exact}
    \boldsymbol{u}(\boldsymbol{x},t) = \frac{1}{2\pi}\frac{(-\boldsymbol{x}_2,\boldsymbol{x}_1,0)}{\boldsymbol{x}_1^2+\boldsymbol{x}_2^2}\left(1-\exp\left({-\frac{\boldsymbol{x}_1^2+\boldsymbol{x}_2^2}{4\nu t}}\right)\right),\quad \text{in} \ \mathbb{R}^3.
\end{equation}

\subsection{3D random vortex dynamics}
In 3D, we utilize Einstein's notation for simplicity. Here, $\varepsilon^{i j k}$ is the Levi--Civita symbol. 
We utilize $(u,v,w)$ to represent the velocity $\boldsymbol{u}$ in $\mathbb{R}^3$. The diffusion process for the 3D NSE is given in \cite{qian2022random} as follows:
\begin{equation}
     d\boldsymbol{X}_t(\boldsymbol{\xi})=\boldsymbol{u}(\boldsymbol{X}_t(\boldsymbol{\xi}),t)dt+\sqrt{2\nu}d\boldsymbol{B}_t,
\end{equation}
where $\boldsymbol{B}_t$ denotes the 3D Brownian motion. Then, we can obtain its corresponding probabilistic representation as follows: \cite{qian2022random}
\begin{equation}
\boldsymbol{u}(\boldsymbol{x},t)_i =\int_{\mathbb{R}^3} \mathbb{E}\left[\varepsilon^{i l k} \frac{\boldsymbol{X}_t(\boldsymbol{\xi})_l-\boldsymbol{x}_{l}}{4 \pi\|\boldsymbol{X}_t(\boldsymbol{\xi})-\boldsymbol{x}\|_2^{3}} [\boldsymbol{G}(\boldsymbol{\xi}, t, 0)]_{k,j}\right] \boldsymbol{\omega}(\boldsymbol{\xi},0)_j d \boldsymbol{\xi},       
\end{equation}
where $\boldsymbol{G}(\boldsymbol{\xi}, t, s) \in \mathbb{R}^{3\times3}$ is a symmetric matrix. The evolution of $\boldsymbol{G}$ obeys the following dynamic system:
\begin{multline}
 \frac{\mathrm{d}}{\mathrm{d} s} [\boldsymbol{G}(\boldsymbol{x}, t, s)]_{i,j}=-[\boldsymbol{G}(\boldsymbol{x}, t, s)]_{i,p} \cdot \\ 
\int_{\mathbb{R}^3} \mathbb{E}\left[H_{j, \alpha}^{p}\left(\boldsymbol{X}_s(\boldsymbol{x})-\boldsymbol{X}_s(\boldsymbol{\xi})\right) [\boldsymbol{G}(\boldsymbol{\xi}, s, 0)]_{\alpha,\beta}\right]\boldsymbol{\omega}(\boldsymbol{\xi},0)_{\beta}\mathrm{d}\boldsymbol{\xi},
\end{multline}
where $ [\boldsymbol{G}(\boldsymbol{\xi}, t, t)]_{i, j}=\delta(i, j)$ and
\[ 
H_{j, i}^{k}(\boldsymbol{x}):=\frac{3}{2} \frac{\boldsymbol{x}_{l}}{4\pi\|\boldsymbol{x}\|_2^{5}}\left(\varepsilon^{k l i} \boldsymbol{x}_{j}+\varepsilon^{j l i} \boldsymbol{x}_{k}\right).
\]
Compared to the 2D cases, calculating the velocity is more complex due to the existence of $\boldsymbol{G}$. DRVN was implemented in 3D as shown in Algorithm~\ref{algo2}.

\begin{algorithm}[H]
\caption{3D Deep Random Vortex Network (DRVN)}
\label{algo2}
\LinesNumbered 
\KwIn{Coordinates $\{\boldsymbol{\xi}^{(i)}\}_{i=1}^{I}$, initial vortex $\{w(\boldsymbol{\xi}^{(i)},0)\}_{i=1}^{I}$, neural network $\{\boldsymbol{u}_{\text{NN}}(\boldsymbol{x},t_m)\}_{m=1}^{M}$.}
Simultaneously initialize the parameter $\{\boldsymbol{\Theta}_{t_m}\}_{m=1}^M$ of the neural networks $\{\boldsymbol{u}_{\text{NN}}(\boldsymbol{x},t_m)\}_{m=1}^M$ via Xavier method\;
Initial $\Bar{G}^n(\boldsymbol{\xi}^{(i)},t_m,t_{m})=\boldsymbol{I}_3$ and $\Bar{\boldsymbol{X}}^n_{t_m}(\boldsymbol{\xi}^{(i)}) = \boldsymbol{\xi}^{(i)}$ for all $m$\;
\For{$E$ epochs}{
Initial $\boldsymbol{X}^n_{t_0}(\boldsymbol{\xi}^{(i)}) = \boldsymbol{\xi}^{(i)}$ and $\boldsymbol{G}^n(\boldsymbol{\xi}^{(i)},t_m,t_m) =\boldsymbol{I}_3$ for all $m$\;
Sample $\{\boldsymbol{x}^{(b)}\}_{b=1}^{B}$ uniformly in $\Omega$\;
$\mathcal{L}$ = 0\;
\For{$M$ steps}{
    $\boldsymbol{X}^n_{t_{m}}(\boldsymbol{\xi}^{(i)}) = \boldsymbol{X}^n_{t_{m-1}}(\boldsymbol{\xi}^{(i)}) + \boldsymbol{u}_{\text{NN}}( \boldsymbol{X}^n_{t_{m-1}}(\boldsymbol{\xi}^{(i)}), t_{m-1}) \Delta t + \sqrt{2 \nu} \Delta \boldsymbol{B}_{m}$\;}
\For{$m$ from $M-1$ to $0$}{
    $[\boldsymbol{G}^n(\boldsymbol{\xi}^{(i)}, t_m, t_{m^{\prime}})]_{o,j}=[\boldsymbol{G}^n(\boldsymbol{\xi}^{(i)}, t_m, t_{m^{\prime}+1})]_{o,j}+\frac{|\Omega|}{NI}\left( [\boldsymbol{G}^n(\boldsymbol{\xi}^{(i)}, t_m, t_{m^{\prime}+1})]_{o,p} H_{j,\alpha}^{p}(\boldsymbol{X}^n_{t_{m^{\prime}+1}}(\boldsymbol{\xi}^{(i)})-\Bar{\boldsymbol{X}}^{n^{\prime}}_{t_{m^{\prime}+1}}(\boldsymbol{\xi}^{(i)}))\right.\cdot$
    $\left.[{\bar{\boldsymbol{G}}}^{n^{\prime}}(\boldsymbol{\xi}^{(i)}, t_{m^{\prime}+1}, 0)]_{\alpha,\beta}\boldsymbol{\omega}(\boldsymbol{\xi^{(i)}},0)_{\beta}\right)\Delta t$, for all $m^{\prime}<m$ \;}
\For{$M$ steps}{
    $\boldsymbol{\hat{u}}_{\text{NN}}(\boldsymbol{x}^{(b)},t_m)_q =\frac{|\Omega|}{NI}\varepsilon^{q l k} \frac{\boldsymbol{X}_t^n(\boldsymbol{\xi}^{(i)})_l-\boldsymbol{x}_{l}^{(b)}}{4 \pi\|\boldsymbol{X}^n_t(\boldsymbol{\xi}^{(i)})-\boldsymbol{x}^{(b)}\|_2^{3}} [{\boldsymbol{G}}^n(\boldsymbol{\xi}^{(i)}, t_m, 0)]_{k,j} \boldsymbol{\omega}(\boldsymbol{\xi}^{(i)},0)_j$ \;
    $\mathcal{L} = \mathcal{L} + \sum_{b=1}^{B} \|\boldsymbol{u}_{\text{NN}}(\boldsymbol{x}^{(b)}, t_m)- \boldsymbol{\hat{u}}_{\text{NN}}(\boldsymbol{x}^{(b)},t_m)\|_2^2$\;
}
Update $\Bar{\boldsymbol{X}}^n_{t_m}(\boldsymbol{\xi}^{(i)}) = {\boldsymbol{X}}^n_{t_m}(\boldsymbol{\xi}^{(i)})$ and  $\Bar{G}^n(\boldsymbol{\xi}^{(i)},t_m,t_{m^{\prime}})={G}^n(\boldsymbol{\xi}^{(i)},t_m,t_{m^{\prime}})$ for all $m$ and $m^{\prime}$\;
Update $\boldsymbol{u}_{\text{NN}}$'s parameters: $\boldsymbol{\Theta}_{t_m} =  \text{optim.Adam}(\boldsymbol{\Theta}_{t_m}, \nabla_{\boldsymbol{\Theta}_{t_m}} \mathcal{L});$ for $m=1,\cdots,M.$
}
\end{algorithm}

\subsection{Problems setups}
In this part, we aim to solve forward problems for a 3D Lamb--Oseen vortex. The vorticity field is initialized as $\boldsymbol{\xi}^{(i)} = (0,0,i/2)$ with $\boldsymbol{\omega}^{(i)} =  (0,0,0.5)$ for $-20\leq i \leq 20$. The Reynolds number is fixed as 2. When training the neural network, we sample data points in the domain $[-2,2]\times[-2,2]\times[-10,10]$ to guarantee the neural networks can learn the information around the initial coordinates $\boldsymbol{\xi}^{(i)}$.

\subsection{Results: Forward problems}
\begin{figure}[!htbp]
\begin{center}
\centerline{\includegraphics[width=0.77\linewidth]{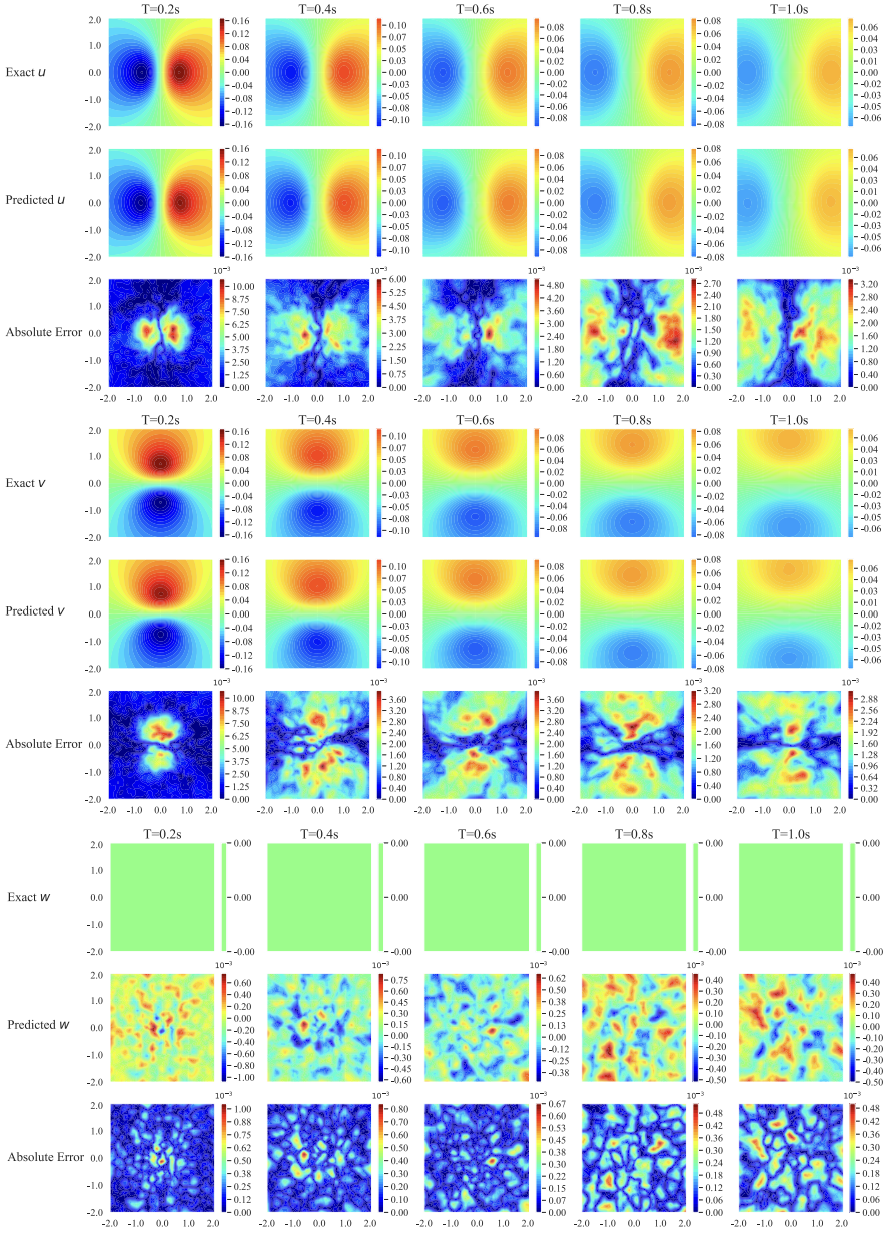}}
\caption{Comparison of the velocity fields between the exact solution and DRVN from $T=0.2$ to $1.0$~s in the periodic domain $[-2,2]^3$ for a 3D Lamb--Oseen vortex.}\label{3d_forward}
\end{center}
\end{figure}

Figure~\ref{3d_forward} shows snapshots of the learned velocity fields and corresponding absolute error during $T \in [0,1]$. Furthermore, we show the effects of the size of the temporal lattice on the relative $\ell_2$ error in Table~\ref{3d_error}. Due to the singular initialization in the 3D Lamb--Oseen vortex, the surfaces of the equations change faster over time. Thus, unlike for the 2D Taylor--Green vortex, the relative errors are more sensitive to the number of temporal intervals.

\begin{table}[!htbp]
\caption{Comparison of the relative $\ell_2$ errors and computational time (per epoch) for different numbers of time intervals $M$ for a 3D Lamb--Oseen vortex.}\label{3d_error}
\centering
\begin{tabular}{rrrr}
\hline
$M$ & ${E}_{T}$ (\%) & ${E}_{[0,T]}$ (\%) & Time (s) \\
\hline
5   & $9.21 \pm0.07$ & $21.71 \pm0.04$ & $0.014$ \\
10  & $3.37 \pm0.09$ & $12.69 \pm0.04$ & $0.028$ \\
20  & $1.58 \pm0.04$ &  $6.83 \pm0.04$ & $0.052$ \\
40  & $1.87 \pm0.07$ &  $4.01 \pm0.05$ & $0.113$ \\
100 & $2.38 \pm0.09$ &  $2.93 \pm0.04$ & $0.278$ \\ \hline
\end{tabular}
\end{table}

\clearpage
\bibliographystyle{plain}
\bibliography{drvn}

\end{document}